\documentclass[fleqn,usenatbib]{mnras}

\usepackage{txfonts}
\usepackage{soul}

\usepackage[T1]{fontenc}

\DeclareRobustCommand{\VAN}[3]{#2}
\let\VANthebibliography\thebibliography
\def\thebibliography{\DeclareRobustCommand{\VAN}[3]{##3}\VANthebibliography}

\usepackage{graphicx}	
\usepackage{amssymb}	
\usepackage{hyperref}
\usepackage{longtable}
\usepackage{tabularx}
\usepackage{afterpage}

\title[Unveiling short period binaries]{Unveiling short period binaries in the inner VVV bulge}

\author[E. Botan et al.]{
E. Botan$^{1,2}$\thanks{E-mail: evertonbotan@ufmt.br},
R. K. Saito$^{1}$,
D. Minniti$^{3,4}$,
A. Kanaan$^{1}$,
R. Contreras Ramos$^{5,6}$,
\newauthor
\ T. S. Ferreira$^{1}$,
L. V. Gramajo$^{7,8}$,
M. G. Navarro$^{3,9,5}$
\\
$^{1}$Departamento de Física, Universidade Federal de Santa Catarina, Trindade, 88040-900, Florianópolis, SC, Brazil.\\
$^{2}$Instituto de Ciências Naturais, Humanas e Sociais, Universidade Federal de Mato Grosso, Res. Cidade Jardim, 78550-728, Sinop, MT, Brazil.\\
$^{3}$Departamento de Ciências Físicas, Faculdade de Ciências Exatas, Universidade Andrés Bello, Av. Fernandez Concha 700, Las Condes, Santiago, Chile. \\
$^{4}$Vatican Observatory, V00120 Vatican City State, Italy. \\
$^{5}$Instituto Milenio de Astrofísica, Santiago, Chile. \\
$^{6}$Instituto de Astrofisica, Pontificia Universidad Catolica de Chile, Vicuna Mackenna 4860, Macul, Santiago, Chile. \\
$^{7}$Universidade Nacional de Córdoba. Observatorio Astronómico de Córdoba, Córdoba, Argentina.\\
$^{8}$Consejo Nacional de Investigaciones Científicas y Técnicas (CONICET), Godoy Cruz 2290, Buenos Aires, CPC 1425FQB, Argentina.\\
$^{9}$Dipartimento di Fisica, Università degli Studi di Roma 'La Sapienza´, P.le Aldo Moro, 2, I-00185 Rome, Italy.\\
}

\date{Accepted XXX. Received YYY; in original form ZZZ}

\pubyear{2020}

\begin{document}
\label{firstpage}
\pagerange{\pageref{firstpage}--\pageref{lastpage}}
\maketitle

\begin{abstract}
Most of our knowledge about the structure of the Milky Way has come from the study of variable stars. Among the variables, mimicking the periodic variation of pulsating stars, are the eclipsing binaries. These stars are important in astrophysics because they allow us to directly measure radii and masses of the components, as well as the distance to the system, thus being useful in studies of Galactic structure alongside pulsating RR Lyrae and Cepheids. Using the distinguishing features of their light curves, one can identify them using a semi-automated process. In this work, we present a strategy to search for eclipsing variables in the inner VVV bulge across an area of 13.4 sq. deg. within $1.68^{\rm o}<l<7.53^{\rm o}$ and $-3.73^{\rm o}<b<-1.44^{\rm o}$, corresponding to the VVV tiles b293 to b296 and b307 to b310. We accurately classify 212 previously unknown eclipsing binaries, including six very reddened sources. The preliminary analysis suggests these eclipsing binaries are located in the most obscured regions of the foreground disk and bulge of the Galaxy. This search is therefore complementary to other variable stars searches carried out at optical wavelengths.
\end{abstract}




\section{Introduction}
Most of our knowledge about the structure of the Milky Way has come from the study of variable stars. Since the "Great Debate" between Curtis and Shapley \citep{Hoskin_1976} and the use of pulsating variable stars as standard candles by Leavitt and Hubble to scale the Galaxy -- and even the Universe -- these stars became important tools in astrophysics \citep{1982eua..book.....S}.  

Chief among them are the RR Lyrae and Cepheid variables, presenting 0.2 to 1.0 day and 1 to 100 days periods, respectively, and accurate period-luminosity relationships \citep[e.g.,][]{Catelan2004,Sandage2009,Turner2010}. Large synoptic surveys of the Milky Way such as OGLE, MOA and SuperWASP in the optical \citep{2003AcA....53..291U, Sumi_2003, 2006PASP..118.1407P}, and the VVV Survey in the near-IR \citep{2010NewA...15..433M} have contributed to the discovery of a large number of pulsating variables, unveiling even the innermost regions of the Galaxy.

Mimicking the periodic variation of pulsating variables are the eclipsing binaries. Most stars in the Milky Way (MW) are known to be in binary systems, with high-inclination systems presenting periodic eclipses. Eclipsing binaries are important in astrophysics because they allow direct measurements of radii and masses of the components, as well as the distance to the system, thus being useful in studies of Galactic structure alongside pulsating RR Lyrae, Cepheids and red clump giants \citep[e.g.,][]{Southworth2005, Helminia2013, Pietrukowicz2020}.

While searching for pulsating variables, a large number of eclipsing binaries have been detected in our searches. In particular the so-called short-period eclipsing binaries (hereafter EBs), presenting periods shorter than 1 day. Since they present distinguishing features in the light curves, algorithms can be trained to select these objects automatically. Here we present the result of a systematic search for EBs in the inner VVV bulge area. The current paper is organised as follows: in Section \ref{sec:vvv_data} we present a description of the VVV time series data and the tiles analysed. In Section \ref{sec:target_sel} we describe the variability indicator used to select the variables. We also describe the implementation of period identification algorithms and the seasonal aliases caused by VVV sampling. In Section \ref{sec:EB_class} we present the techniques used to identify the EBs candidates. In Section \ref{sec:discussion} we compare our sample with catalogues of variables stars, leading us to the conclusions presented in Section~\ref{sec:conclusion}.

\begin{figure*}
	\includegraphics[width=0.8\textwidth]{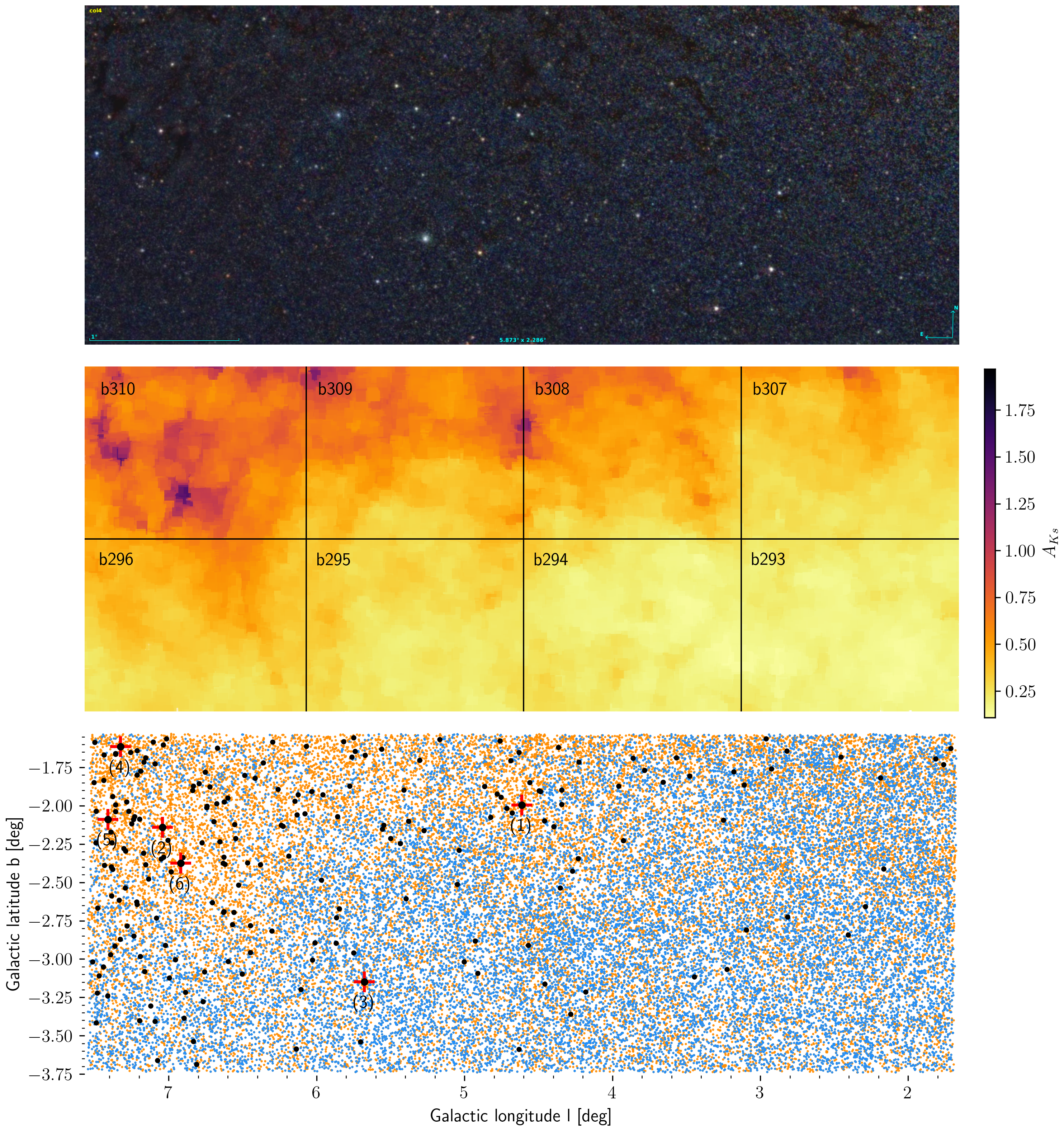}
    \caption{The top image is the VVV $JHK_{\rm s}$ false-colour image corresponding to the region covered by the VVV tiles b293 to b296 and b307 to b310. The image is oriented in Galactic coordinates within $1.68^{\rm o}<l<7.53^{\rm o}$ and $-3.73^{\rm o}<b<-1.44^{\rm o}$, resulting in a total area of $\sim 13.4$ sq. deg. (see Section 2). The middle image is the region extinction map constructed using BEAM calculator  \citep{Gonzalez2012}. The bottom image is the spatial distribution of all potential VVV variables (orange) and VSX and OGLE-IV matches (blue). The black dots are our new EBs and the targeted and numbered are the most reddened ones.}
    \label{fig:fov}
\end{figure*}

\section{Near-IR VVV Data} \label{sec:vvv_data}

The VISTA Variables in V\'{\i}a L\'actea (VVV) is a European Southern Observatory (ESO) public survey which has recently completed near-infrared (near-IR) observations  of 562~sq.~deg.  area of the Galactic bulge and the southern disk. The VVV observational strategy consisted of two sets of multicolour $ZYJHK_{\rm s}$ observations taken in 2010 and 2015, plus a variability campaign of typical $60-300$ observations  in $K_{\rm  s}$~band from 2010 to 2016 \citep[][]{2010NewA...15..433M, 2012A&A...537A.107S, Hempel_2014}.

Standard VVV aperture photometry on the stacked VVV tile images for both colour and variability data is provided by the  Cambridge Astronomical Survey  Unit  \citep[CASU;  for  details  see][]{2012A&A...537A.107S} and publicly available through the VISTA Science Archive (VSA\footnote{\url{http://horus.roe.ac.uk/vsa/index.html}}). Following the aperture catalogues, point-spread function (PSF) photometry has also been performed on the VVV images, producing finer catalogues particularly important for the inner regions of our  Galaxy, where the high crowding limits the use of the aperture photometry \citep[e.g.,][]{Mauro_2013,2017A&A...608A.140C,2018A&A...619A...4A, 2019A&A...629A...1S}. 

While most of the $K_{\rm s}$ light curves for the VVV bulge area reach about 100 epochs, a total of eight tiles in the inner bulge have been observed for 300 epochs. These tiles -- namely b293 to b296 and b307 to b310 -- cover a total area of $\sim13.4$ sq. deg. within Galactic coordinates $1.68^{\rm o} <l<7.53^{\rm o}$ and $-3.73^{\rm o}<b<-1.44^{\rm o}$, partially covering the well studied Baade's window (see Fig.~\ref{fig:fov}). PSF photometry on the individual pawprint observations \citep[see][for details]{2017A&A...608A.140C} for these tiles allowed us to extract a total of 11,621,729 light curves that presented at least 25 observations in order to detect RR Lyrae and other pulsating stars efficiently \citep[e.g.,][]{2012sngi.confE..17D}.

\section{Target Selection} \label{sec:target_sel}

When searching for variables, a variability indicator is required. The choice of this indicator depends basically on the investigated period and the sample spacing. When the light curve is well sampled, with several measurements in a pulse cycle, we can use indicators for correlated observations, such as the Welch-Stetson indices \citep{Welch1993, Stetson1996}. However, when sampling is performed with much greater spacing than the stellar period, other indices should be used. In these cases, most of these indices take into account the dispersion of the measures concerning the average of the light curve \citep[for a detailed description of some variability indices for correlated and uncorrelated observations see][]{2016A&A...586A..36F, 2017A&A...604A.121F}. Both types can be used in VVV, but for the range of period of short-period binaries and RR Lyrae the sparse indices are more suitable. Following \citet{Carpenter2001} we used the reduced chi-square in this work:
\begin{equation}
    \chi^2=\frac{1}{n-1} \sum_{i=1} ^n \frac{(m_i - \overline{m})^2}{\sigma_i ^2}
	\label{eq:chi2}
\end{equation}
where \textit{n} is the number of epochs, $m_i$ is $K_{\rm s}$ magnitude measurement at epoch \textit{i}, $\overline{m}$ is the $K_{\rm s}$ measurements mean and $\sigma_i$ is the photometric error at epoch \textit{i}.

\begin{figure}
    \includegraphics[width=\columnwidth]{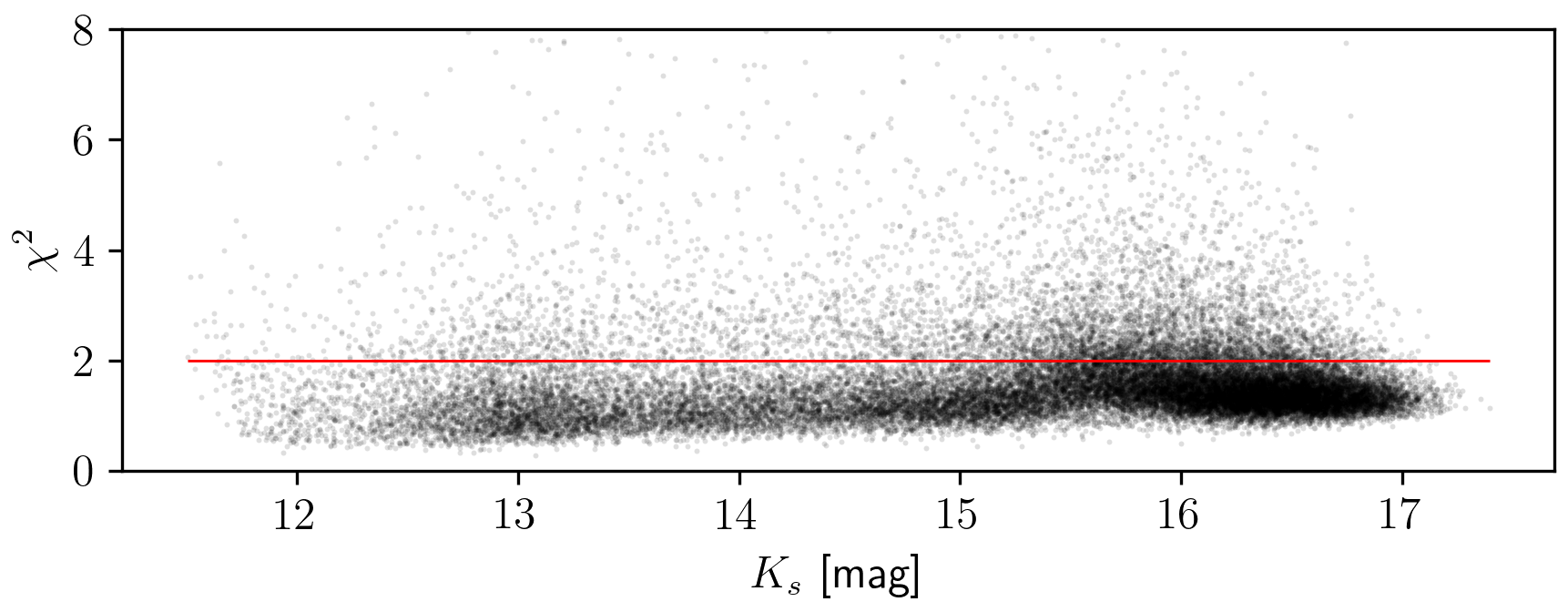}
    \caption{Reduced $\chi^2$ as a function of magnitude from the light curves from a single chip of tile b293. We selected all stars above the red line.}
    \label{fig:candidates}
\end{figure}

Operationally, variable stars exhibit larger photometric variations than expected by random noise throughout a time series, were $\chi^2$ should be close to 1. Fig.~\ref{fig:candidates} shows the reduced $\chi^2$ as function of magnitude for the light curves from a single chip from tile b293. Most of the stars are scattered close to $\chi^2 \sim 1$. Thus, following \citet{Carpenter2001} we selected all sources with $\chi^2 > 2$. As the $\chi^2$ is sensitive to outliers, measurements that showed deviation greater than 3 sigmas were removed from the light curves before passing through the reduced chi-squared test.

These selected candidates comprise 2,950,742 light curves within the $\sim$~13.4 sq. deg. area, including stars from the foreground MW disk and bulge, and even from the halo in the background. There are also six known globular clusters in the region: Djorg 2, Terzan 9, Terzan 10, NGC 6540, NGC 6544 and NGC 6553  \citep[][the Harris catalogue version 2010]{1996AJ....112.1487H}. The candidates were searched for periodicity using Lomb-Scargle \citep[LSG.][]{Lomb1976,Scargle1982} and Phase Dispersion Minimization \citep[PDM.][]{Stellingwerf1978} covering 0.1 to 1000 days. Any period with a false alarm probability  \citep[FAP.][]{Scargle1982,Schwarzenberg_Czerny_1997} smaller than 0.1\% was considered a good detection for both methods. 

To reduce the number of misidentifications and to set the reliability of signal detection we compared the LSG results with PDM, making sure the detected peaks agree within 1\% in the values of their periods. As PDM may find the correct (double found by LSG) period for binaries, we also took into account the harmonics ($P \times n$ and $P/n$ up to $n = 4$ harmonics), which represented approximately 43\% of all candidates. 

\begin{figure}
	\includegraphics[width=\columnwidth]{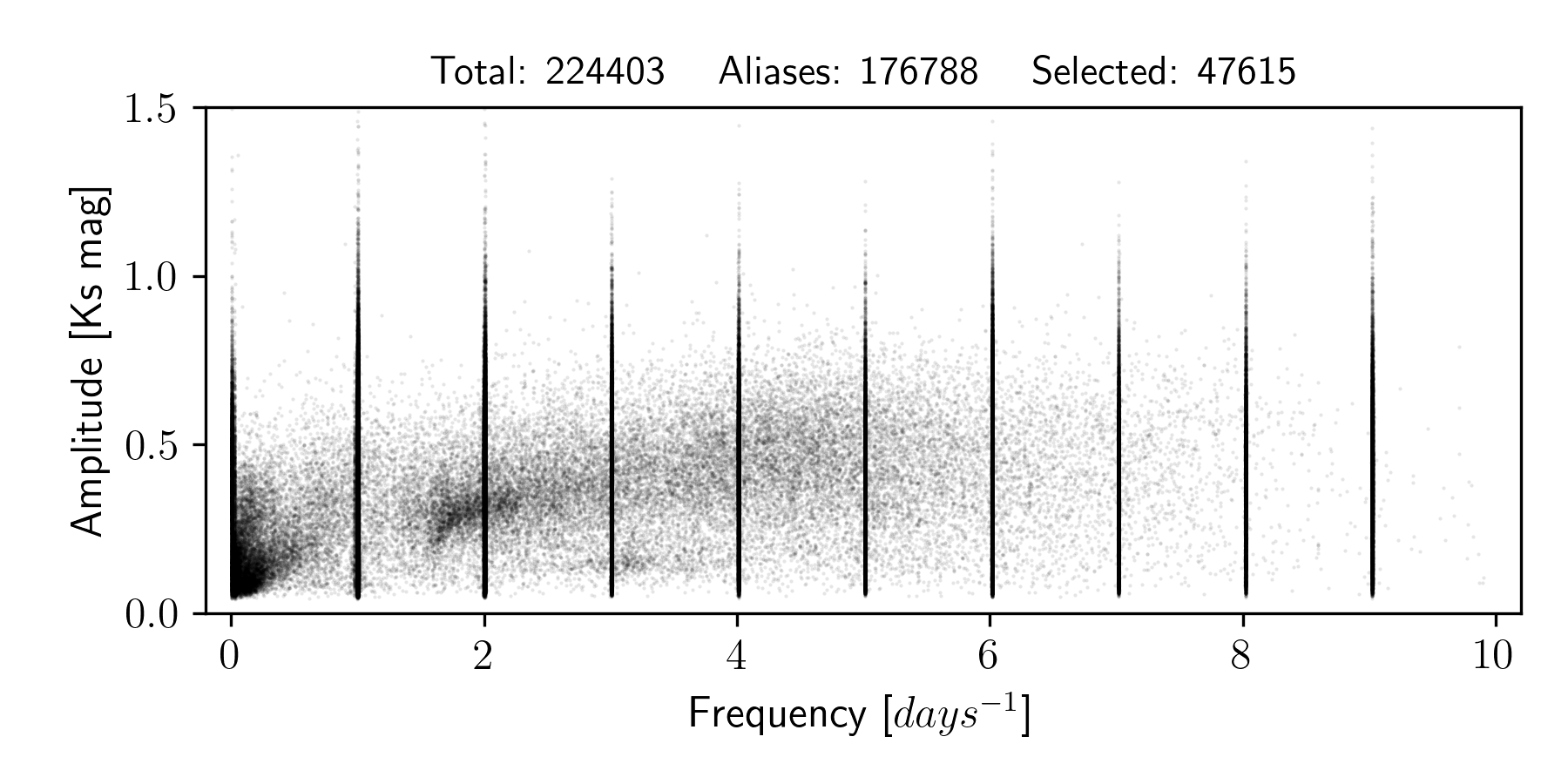}
    \caption{Plot of amplitude versus frequency. The seasonal frequencies are signed by vertically aligned dots at integer multiples of day$^{-1}$. The frequency intervals used to remove the stars over these alias were calculated by finding the mean and the standard deviation close to the harmonics. The locus of RRab type variables at frequency 2 d$^{-1}$ and amplitude 0.25 mag is clearly seen.}
    \label{fig:ampvsfreq}
\end{figure}

\begin{figure}
	\includegraphics[width=\columnwidth]{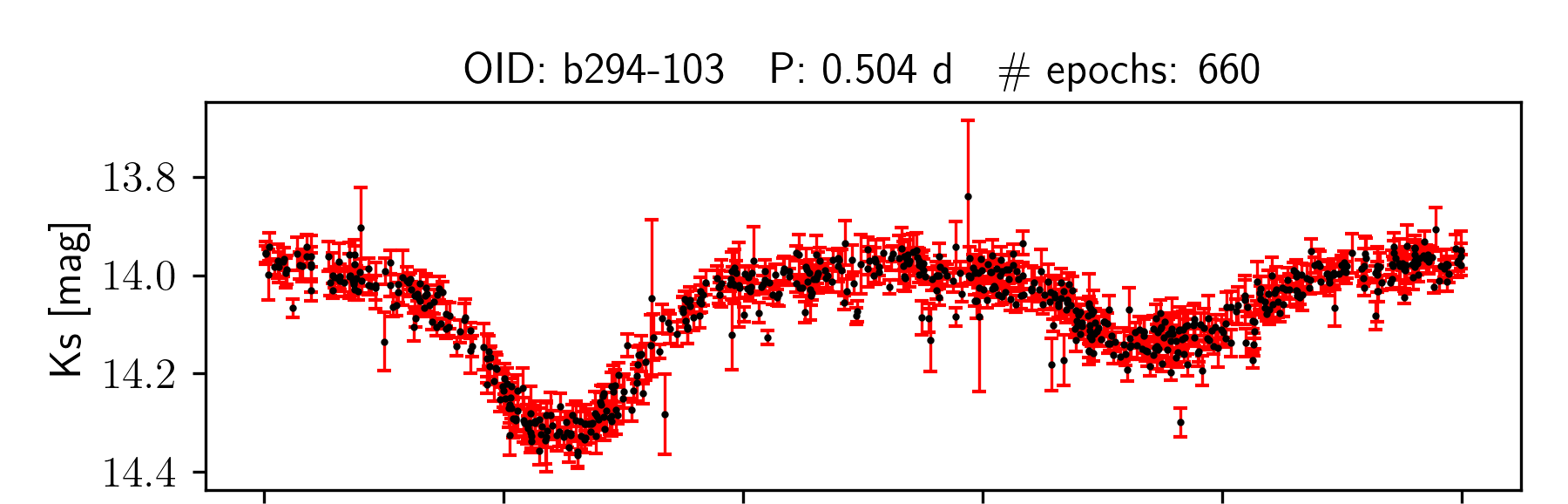}
	\includegraphics[width=\columnwidth]{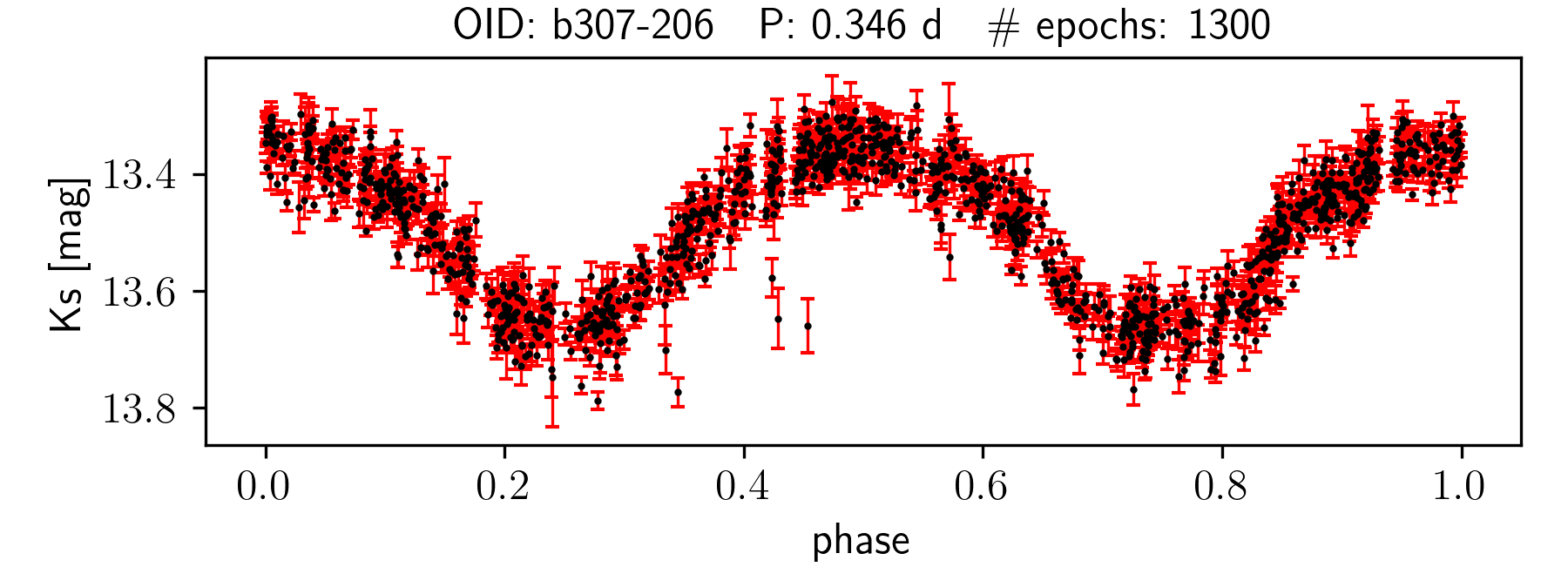}
    \caption{Illustrative examples of light curves for variables on overlapping images from different chips thus presenting twice the number or epochs of the tile. In the top is an example of asymmetric EB while the bottom panel shows a symmetric one.}
    \label{fig:EB_overlap}
\end{figure}

Both LSG and PDM are susceptible to seasonal aliases caused by sampling. Fig.~\ref{fig:ampvsfreq} shows these alias frequencies as vertical columns at integer multiples of day$^{-1}$. Stars whose peaks were identified as aliases have been removed from our analysis, leaving a total of 47,615 variable sources covering the period range of 0.1 $-$ 1000 days. Since this work is focused on short-period variables, in the period range of RR Lyrae stars ($P \lesssim 1$~day), we limited our sample to that extent, remaining with 23,857 variables in the period range from 0.1 to 1 day from the original $\sim 3$ million light curves. Objects with periods longer than 1 day are the subject of a forthcoming paper. From those we have 1,571 variables laying in overlapping regions across the images (from different chips), thus presenting two (or more) sets of measurements that could be combined in light curves with up to 1300 points (see Fig. ~\ref{fig:EB_overlap}).

\section{Eclipsing Binaries Classification} \label{sec:EB_class}

\subsection{Previous works}

It is not a trivial task to classify variable types based solely on the time series. Some efforts on this subject have already been done by \citet{Debosscher_2007} trying to use Fourier coefficients from a sinusoidal fit to the light curves. Frequency range and the first harmonic amplitude are often used to tell Miras from RR Lyrae and Cepheids, however, they are not efficient to separate EBs from RR Lyrae \citep[see Fig. 2 from][for details]{Debosscher_2007}.

\citet{Sumi_2004}, based the \citet{Alard_1996} method, tried to separate RR Lyrae from other variables by 1) fitting an ellipse to the locus of stars in a diagram representing the ratio of second to first harmonic amplitude ($R_{21} = A_2/A_1 $) and the phase difference between these harmonics ($\phi_{21} = \phi_2 - 2\phi_1$) \citep[see Fig. 11 from][for more details]{Sumi_2004}; 2) visual inspection of the remaining light curves to remove contamination and any lasting variables.

Other approaches are currently being developed for the VVV data using machine learning algorithms, such as the VVV Templates Project \citep{2014A&A...567A.100A} and the method of \citet{Medina_2018}. In both cases an automated classification of variable sources is performed based on the comparison of VVV light curves with high-quality near-IR light curves of known variable stars. However, the most common way to classify variable stars is through visual inspection of their light curves as done by \citet{Montenegro_2019} and \citet{2020PASA...37...54G} which identified the first EBs from the VVV survey. In this work, we seek a different approach, by considering the asymmetry of eclipsing binary light curves to unveil short period binaries in the inner VVV bulge.

\subsection{Identifying asymmetric EBs}
\label{sec:assymetric_EBs}

\begin{figure}
	\includegraphics[width=\columnwidth]{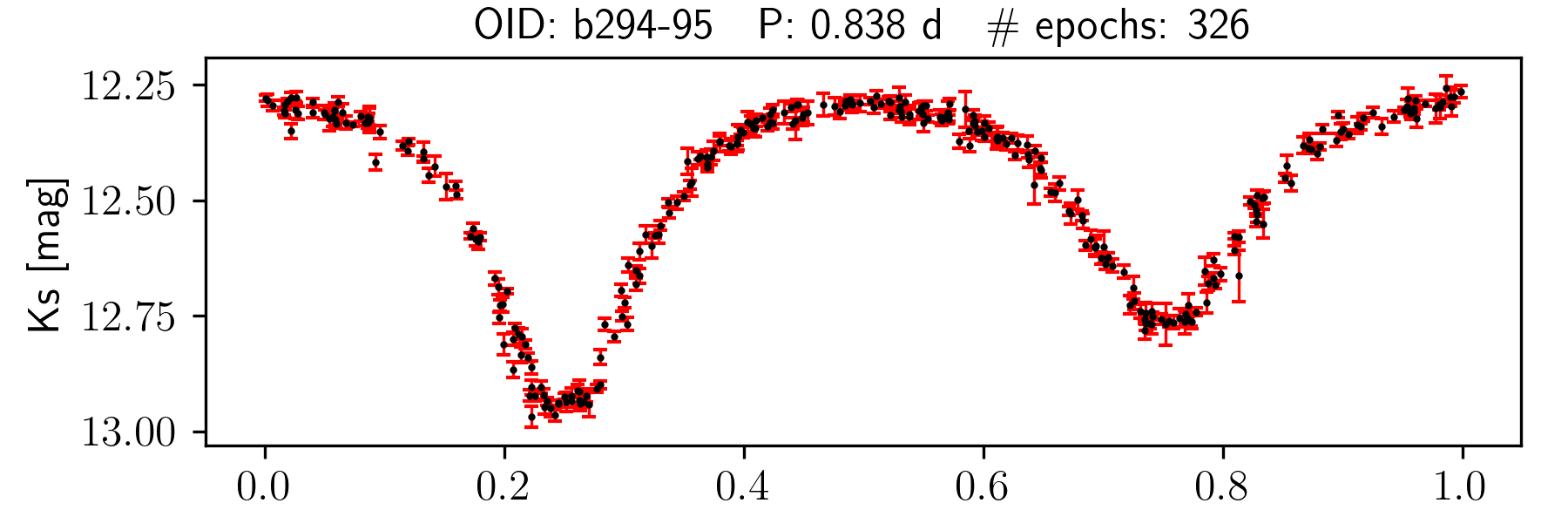}
	\includegraphics[width=\columnwidth]{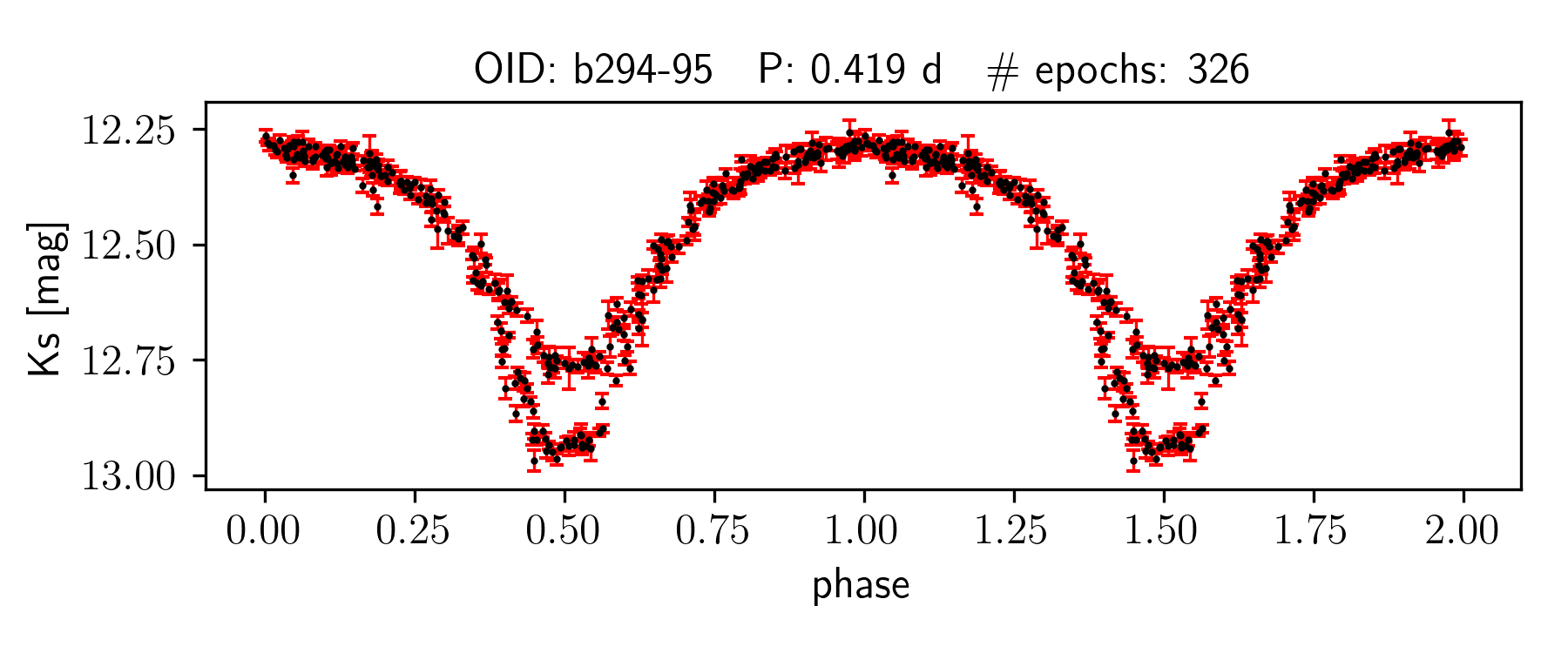}
    \caption{Phase-folded light curve for an EB with twice the LSG period (top) and the LSG period itself (bottom).}   
    \label{fig:2P_lc}
\end{figure}

The EBs are classified mainly into three classes: EA (Algols), EB ($\beta$ Lyrae) and  EW (W Ursae Majoris). The EA and EB are known to present eclipses with a large difference between the depth of the two minimums, as shown in the first plot of Fig.~\ref{fig:2P_lc}. EWs on the other hand have symmetrical light curves \citep{Kallrath_2009,Catelan2015book}.

Despite the asymmetry of the EA and EB light curves, LSG generally identifies half of the true period for these variables, and a plot of the same light curve with the LSG period returns a completely different shape, as seen in the second plot of Fig.~\ref{fig:2P_lc}. To overcome this we classify the asymmetric EBs by analysing the depth of their two valleys on plots with twice the LSG period.

To compare the two minimums we performed a Fourier fit of $N = 5$ harmonics (to avoid over-fitting) for each light curve in phase space, given its period ($P$, twice the LSG period) and average magnitude $\langle m \rangle$, using the Fourier series $f(t)$: 
\begin{equation}
    f(t) = \langle m \rangle + \sum_{n=1} ^N A_n sin \left( \frac{2\pi n t}{P} + \phi_n \right)
	\label{eq:F_Sin}
\end{equation}
where the amplitude $A_n$ and phase $\phi_n$ for each $n$ harmonic is calculated.

\begin{figure}
	\includegraphics[width=\columnwidth]{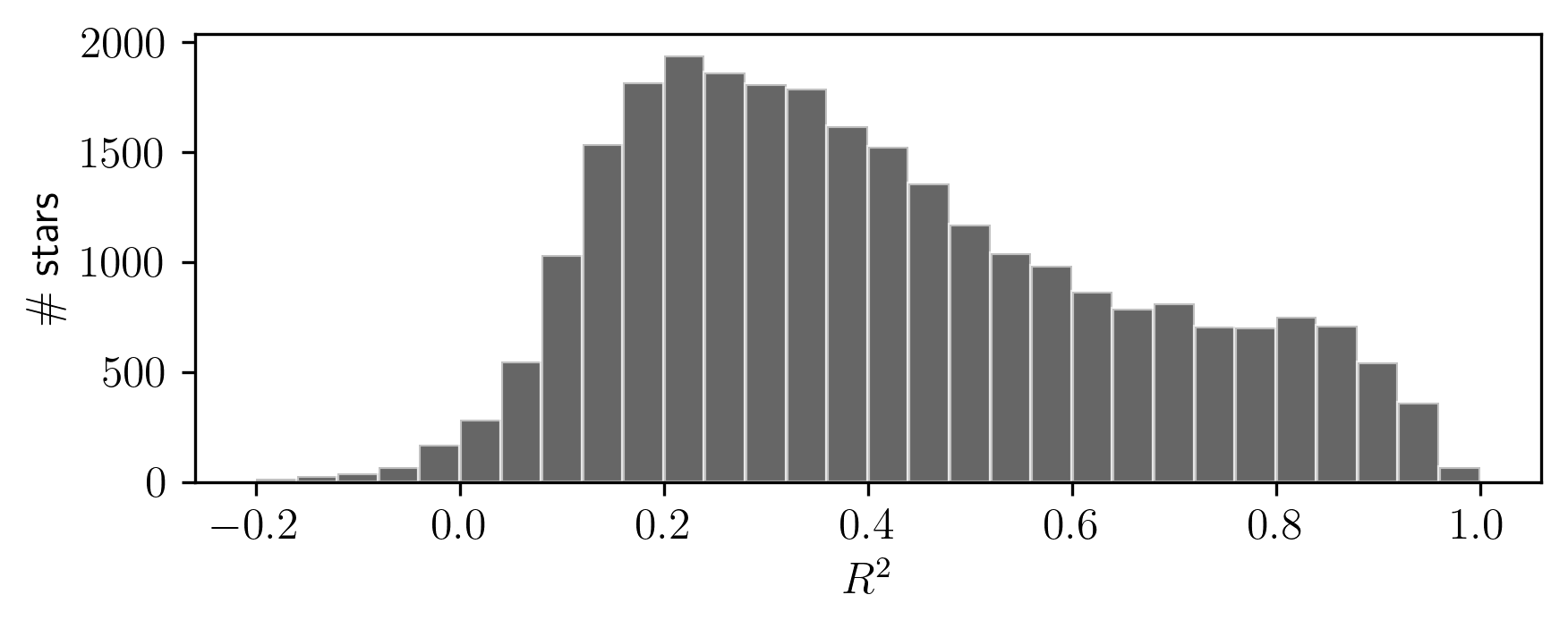}
    \caption{$R^2$ distribution from the Fourier fits to our sample.}   
    \label{fig:R2_dist}
\end{figure}

We implemented the $R^2$ statistics parameter as a selection criterion to our variables since it shows how close the data are to the fitted model. Fig.~\ref{fig:R2_dist} shows the $R^2$ distribution from the Fourier fits to our sample of 23,857 variables. The distribution peaks around $R^2 = 0.2$ and then starts to flatten after $R^2 \sim 0.6$. Therefore, to select the smoothest light curves, we used $R^2_{\rm index} = 0.6$ as a threshold limit, remaining a total of 6,308 objects with $R^2 > 0.6$. The fitted curve, then, allows us to characterise the minima of the eclipses. By comparing the depths of the valleys with the data dispersion about the fit, one can infer if the difference between the valleys is larger than the standard deviation of the residual, and then classify the star as an asymmetric binary. This procedure allowed us to automatically select 850 bonafide asymmetric EBs.

We notice that the $R^2$ does not indicate whether a regression model is adequate, however, the Fourier series seems to be a good model for most variables, and it has been used in many similar searches for periodic variables \citep[e.g.][]{Debosscher_2007, Medina_2018}.

\subsection{Symmetric EBs in the sample}
\label{sec:symmetric_EBs}

As expected, a large number of symmetric EBs should also be present in our sample. For the sake of completeness, a visual inspection on all $R^2 > 0.6$ symmetric light curves was also performed to distinguish symmetric EBs from RR Lyrae and other contaminators. This procedure included 1,840 potential EBs in our final sample, resulting in a final sample of 2,690 sources, including 261 EBs in overlapping regions. Illustrative examples of light curves from symmetric and asymmetric EBs laying in overlapping regions of different chips are presented in Fig.~\ref{fig:EB_overlap}.

\begin{figure}
	\includegraphics[width=\columnwidth]{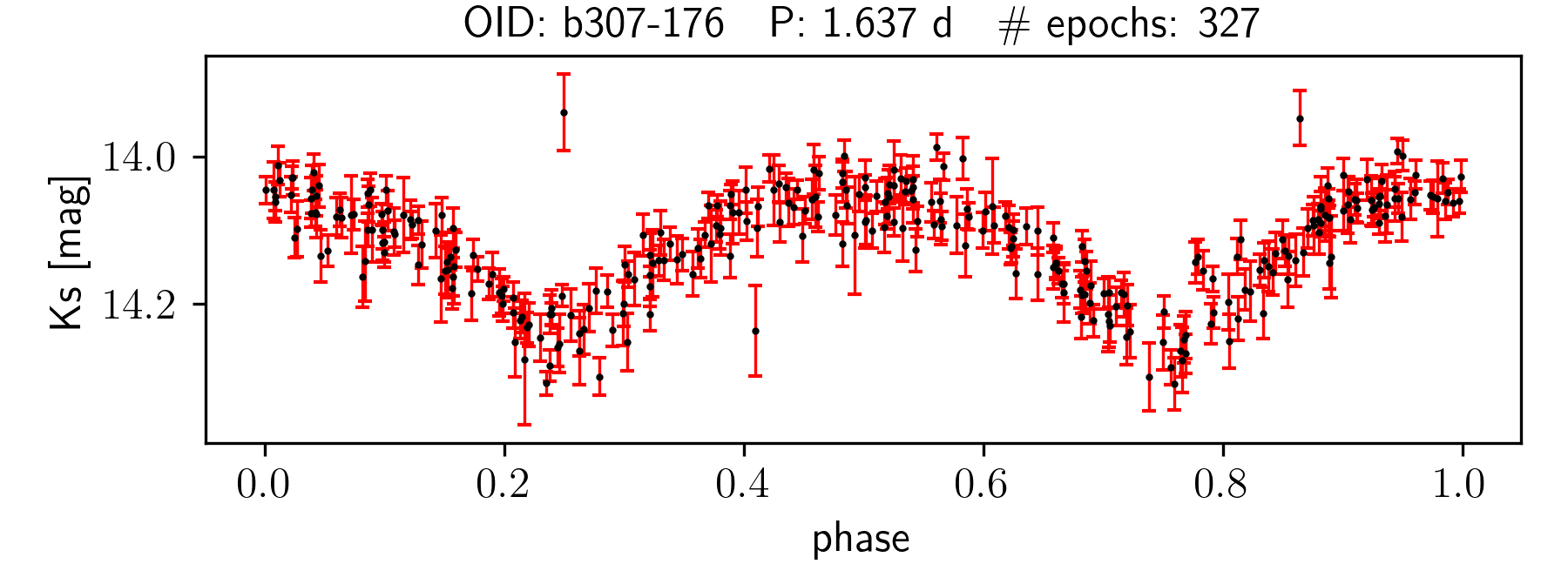}
    \caption{Example of light curve with previous classification of RR Lyrae selected as symmetric EB in this work (see Section 4.3).}
    \label{fig:EB_mis}
\end{figure}

\begin{figure*}
    \flushleft
	\includegraphics[width=\textwidth]{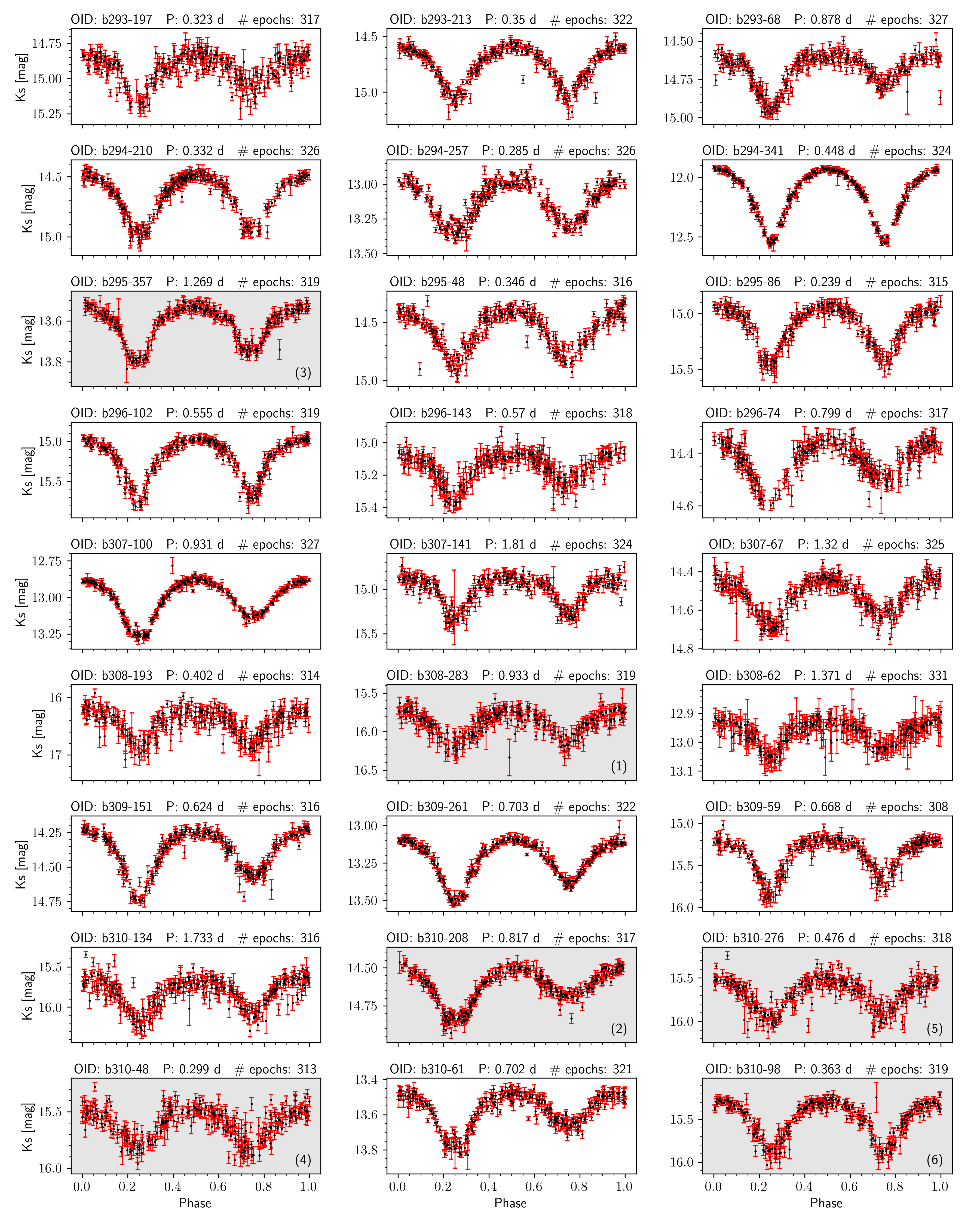}
    \caption{Examples of phase folded light curves for new EBs discovered in each of the eight tiles under analysis in this work. The shaded plots are numbered accordingly to the 6 most reddened EBs signed in Fig. ~\ref{fig:cmd}.}
    \label{fig:new_EBs}
\end{figure*}


\section{Comparison with OGLE, VSX, VVV-CVSC and GAIA} \label{sec:discussion}
\label{sec:cat}

We compared our resulting sample with OGLE-IV\footnote{\url{http://ogle.astrouw.edu.pl/}}, VSX\footnote{\url{https://www.aavso.org/vsx/}}  and VVV-CVSC\footnote{\url{http://horus.roe.ac.uk/vsa/vvvGuide.html\#VIVACatalogue}} catalogues. The VSX is expected to have all the OGLE stars, but recent OGLE objects have not been added yet, so we compare them with both catalogues to ensure that we select all known variables within the area. We found a total of 45,790 variables classified as EBs within the period range of 0.1 to 1 day over the VVV area studied here. 

We cross-matched the catalogues using 1 arcsec maximum separation as a matching criterion. As result, we found that 2,478 of 2,690 EBs in our sample are previously known variables, most of them classified also as eclipsing binaries. From those, just a small number of 30 objects have a previous classification as RR Lyrae. These RR Lyrae were included in our sample during the visual inspection of symmetric light curves because of the difficulty in distinguishing them from EBs in the VVV $K_{\rm s}$ band data (see Fig.~\ref{fig:EB_mis}).

The VVV-CVSC \citep{Ferreira_2020} is a recent catalogue of unclassified variables from all VVV survey area. A cross match revealed 18,787 sources from our sample of 47,615 potential variables. This difference is mainly caused by the selection criteria used on the variability indicator. They used the deviation grater than $1.5\sigma$ from the expected light curve RMS in order to balance true variables and contamination. However, as pointed out by the authors, it may miss small amplitude variables as well. From our EB candidates 2.230 were selected by VVV-CVSC as unclassified variables, including 173 of the 212 new EBs.

Thus, with this method we identify 212 new EBs across eight inner VVV tiles covering an area of $\sim 13.4$~sq. deg. Light curve examples of new EBs are presented in Fig.~\ref{fig:new_EBs} while in Appendix A we present a table with coordinates, periods, and near-IR magnitudes and colours for all the new EBs. While these 212 new variables represent an increment of $\sim 1\%$ in the total number of previously known objects in the area, we highlight that all new objects have been selected over a conservative threshold ($R^2~{\rm index} > 0.6$) in this first approach, thus presenting a sample of high-quality light curves for bonafide variables. Moreover, as we will discuss next (e.g., see Table 1), the number of new objects per tile varies by a factor $> 20$, where most of the sources lay towards higher extinction regions. The spatial distribution of the new EBs is presented at the bottom panel of Fig.~\ref{fig:fov} (black dots) while the distribution per tile is presented in Table~\ref{tab:EB_found}, which includes also the previously known objects in our sample. While previously known EBs are evenly distributed for new objects a major contribution is made in tiles b310, b309, b308 and b296 (top-left corner of the image in Fig. \ref{fig:fov}) where the extinction is severe and optical surveys can not detect variable stars.

\begin{table}
	\centering
	\caption{The EBs found by tile, oriented from top left of the bottom image of Fig.~\ref{fig:fov}.}
	\label{tab:EB_found}
	\begin{tabular}{lcccc} 
		\hline
		 & b310 & b309 & b308 & b307  \\
		\hline
		New EBs & 84  & 32  & 22  & 10  \\
		Previously known   & 231 & 299 & 295 & 280 \\
		Totals    & 315 & 331 & 317 & 290 \\
		\hline
	\end{tabular}
	
		\begin{tabular}{lcccc} 
		 &  b296 & b295 & b294 & b293 \\
		\hline
		New EBs & 42  & 12  & 6   & 4   \\
	    Previously known  & 325 & 394 & 358 & 296 \\
		Totals   & 367 & 406 & 364 & 300 \\
		\hline
	\end{tabular}
\end{table}

\begin{figure}
    \centering
	\includegraphics[width=\columnwidth]{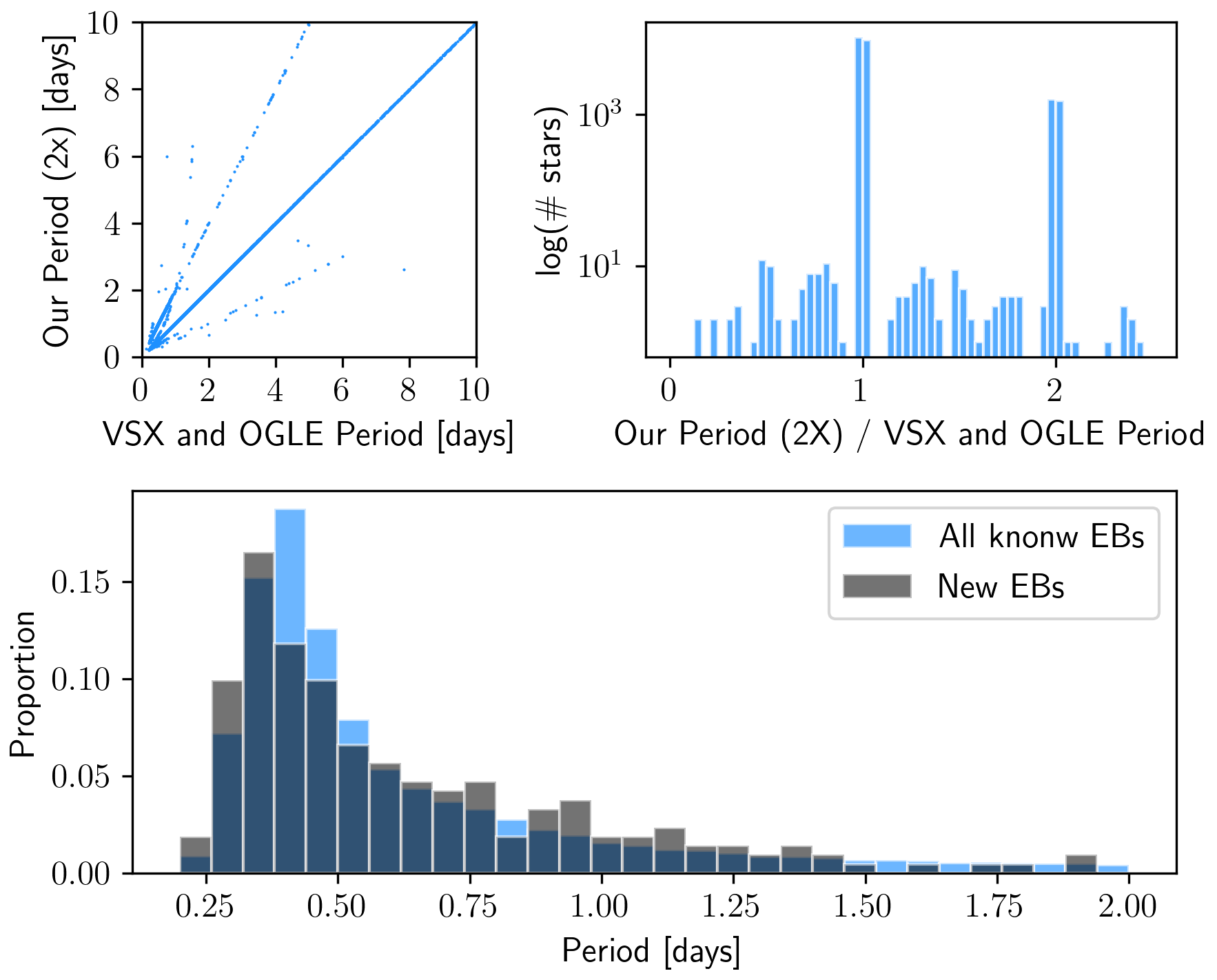}
    \caption{On the top left we have a plot of OGLE and VSX period against twice our LSG periods for all VVV potential variables. On the top right we have the histogram of the ratio between our period (2X) and the matched sources. At the bottom we have the histogram of the periods from all OGLE and VSX EBs within the tiles area and the new EBs found.}
    \label{fig:paccuracy}
\end{figure}

We made use of previously known objects (from OGLE and VSX) to check the accuracy of the periods found by our method. From the original sample containing objects with up to 1000 days period, 85.9\% coincide with known periods with an accuracy of 0.1\%. This comparison is presented in Fig.~\ref{fig:paccuracy}. For some objects there is a resonance in multiples of the period, mainly because we plot twice the LSG period, so that pulsating variable such as RR Lyrae and Cepheids will be shown with their period doubled. Some other misidentification may also be caused by the VVV sampling aliases \citep[see Sec. 7.2 of ][for a detailed discussion]{VanderPlas_2018}. With a proper classification the accuracy should increase, for instance, for our final sample of EBs the agreement is 98.7\%. Thus, we expect a similar reliability for the periods -- as well as for the classification -- of our new EBs. 

The distribution in periods is similar to the all known EBs, however, the periods for the new EBs appear to peak slightly towards shorter periods in relation to all known EBs in the area. This small difference is inconclusive because it could result from the selection criteria. It would be worth to explore these binaries through determination of orbital parameters using simulated models like the PHOEBE model \citep{Pr_a_2005,Andrej2018,Conroy_2020}.

\begin{figure}
    \centering
	\includegraphics[width=\columnwidth]{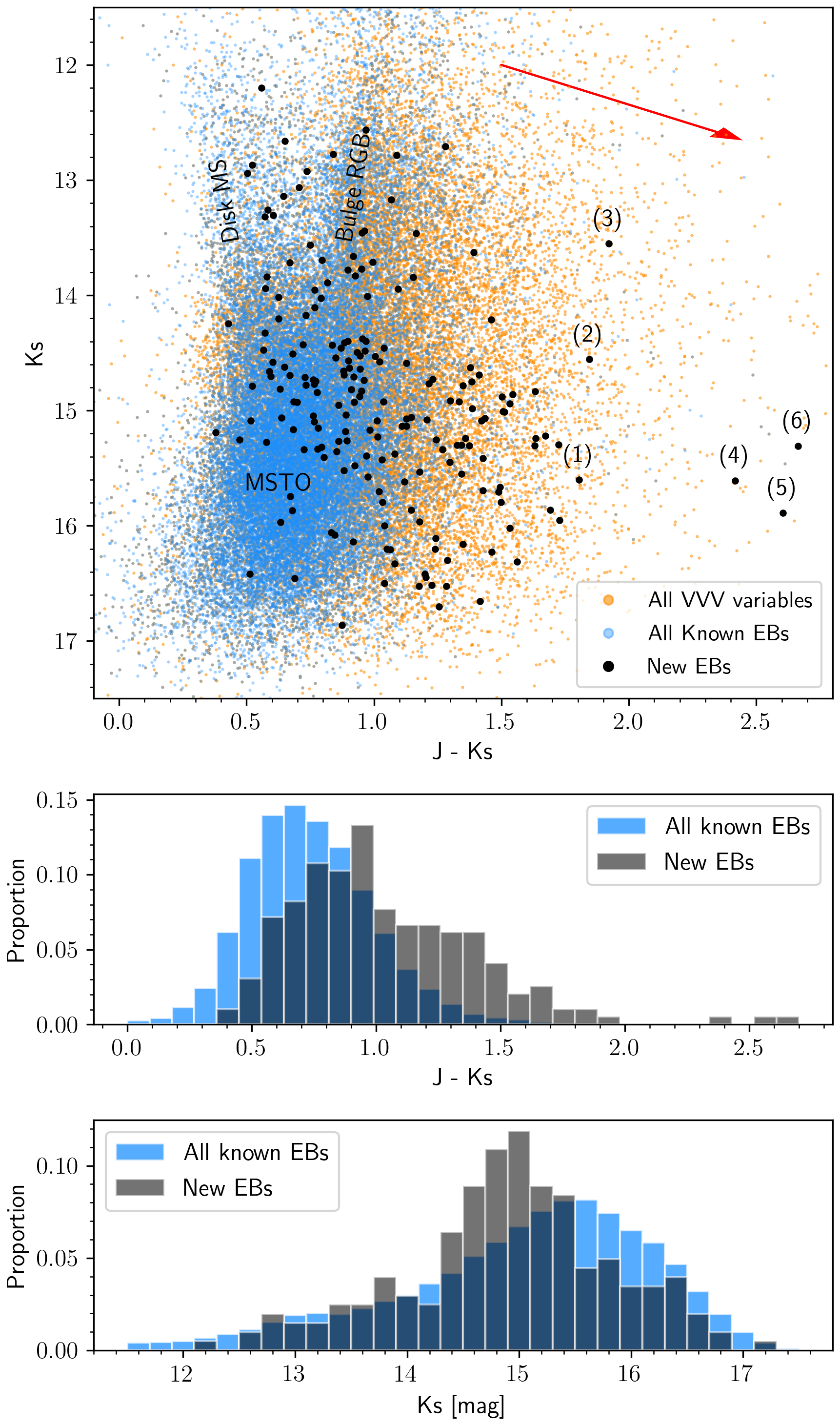}
    \caption{At the top we have a colour magnitude diagram (CMD) for all known EBs (blue) from OGLE and VSX with $J$ and $K_{\rm s}$ mag, VVV sources (orange) and the new EBs (black dots). A reddening vector associated with an extinction of $E(J-K_{\rm s})=0.85$ mag, and assuming the \citet{1989ApJ...345..245C} extinction law, is shown as a red arrow. Disk, Bulge and MSTO sequences are labelled accordingly \citep[e.g., Besançon and Trilegal synthetic models;][respectively]{2003A&A...409..523R,2005A&A...436..895G}. Just 197 of the new EBs have both $J$ and $K_{\rm s}$ bands. The light curves from the numbered EBs are plotted in Fig. ~\ref{fig:new_EBs}. At the middle we have the colour distribution for the new EBs. At the bottom we have the distribution of magnitudes of all known EBs within the tiles area  and the new EBs.}
    \label{fig:cmd}
\end{figure}

Comparing the magnitude distributions (see Fig. ~\ref{fig:cmd}), EBs from OGLE/VSX seem to be equally distributed along the luminosity function while the new EBs seem to present a double peak, at $K_{\rm s} \sim 13.8$~mag and $K_{\rm s} \sim 15.0$~mag. These could be caused either by different extinction levels affecting the sources or by the Milky Way structure, as the presence of the spirals arms in the foreground disk or even from objects belonging to the MW bulge. The CMD shows that new EBs are highly affected by reddening and seems to be distributed along the disk and bulge sequences.  The mean reddening for our sources accordingly with the BEAM\footnote{\url{http://mill.astro.puc.cl/BEAM/calculator.php}} calculator \citep{Gonzalez2012} is $<E(J-K_{\rm s})>\ = 0.85$ mag, with objects reaching up to $E(J-K_{\rm s}) = 2.1$ mag. In comparison, known EBs are less affected by dust and mostly distributed along the disk sequence, including the region of the Main-Sequence Turn-Off (MSTO). The finding of such highly reddened EBs highlights VVV ability to find objects invisible to optical surveys. 

\begin{figure}
    \centering
	\includegraphics[width=\columnwidth]{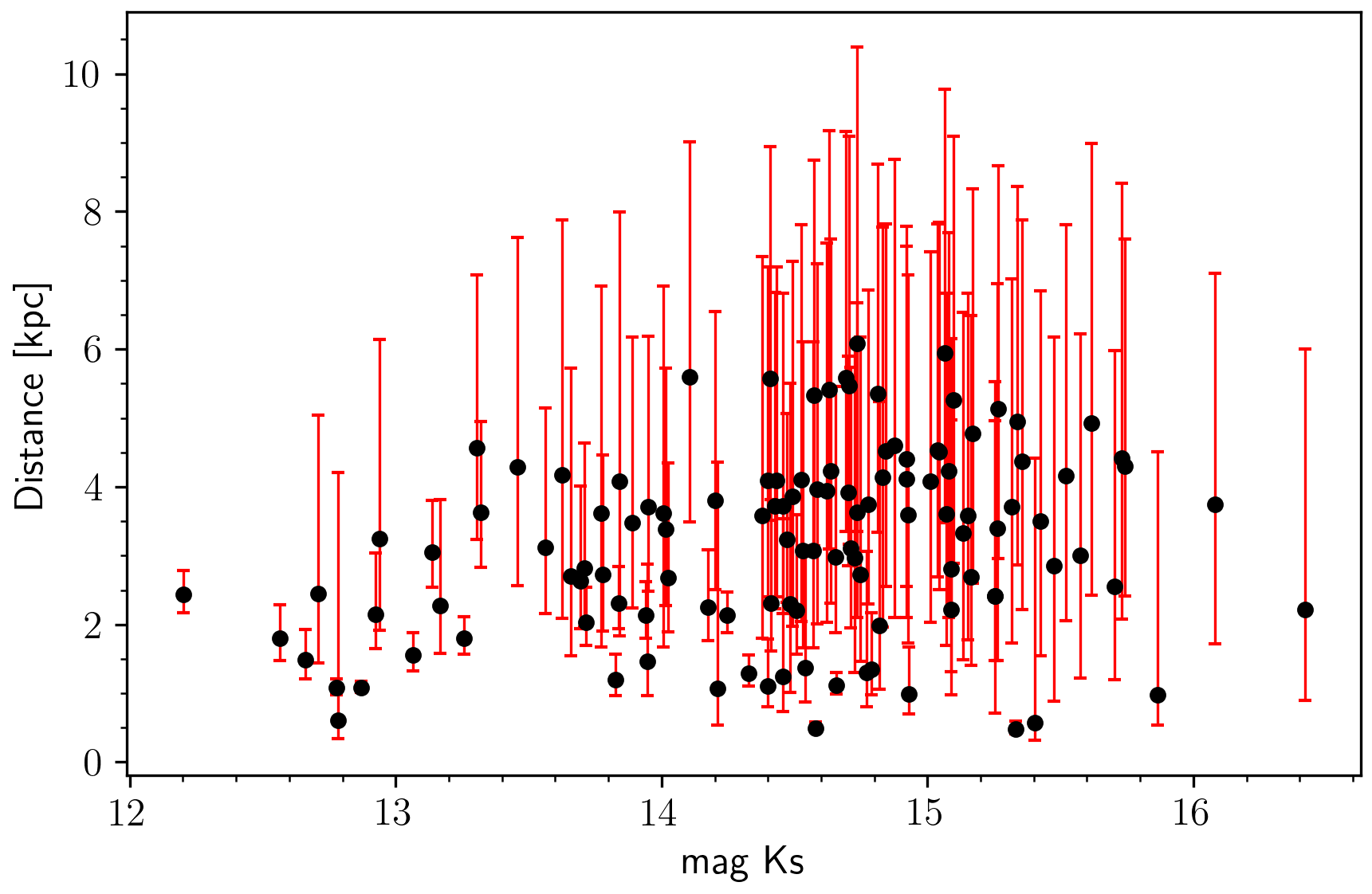}
	\includegraphics[width=\columnwidth]{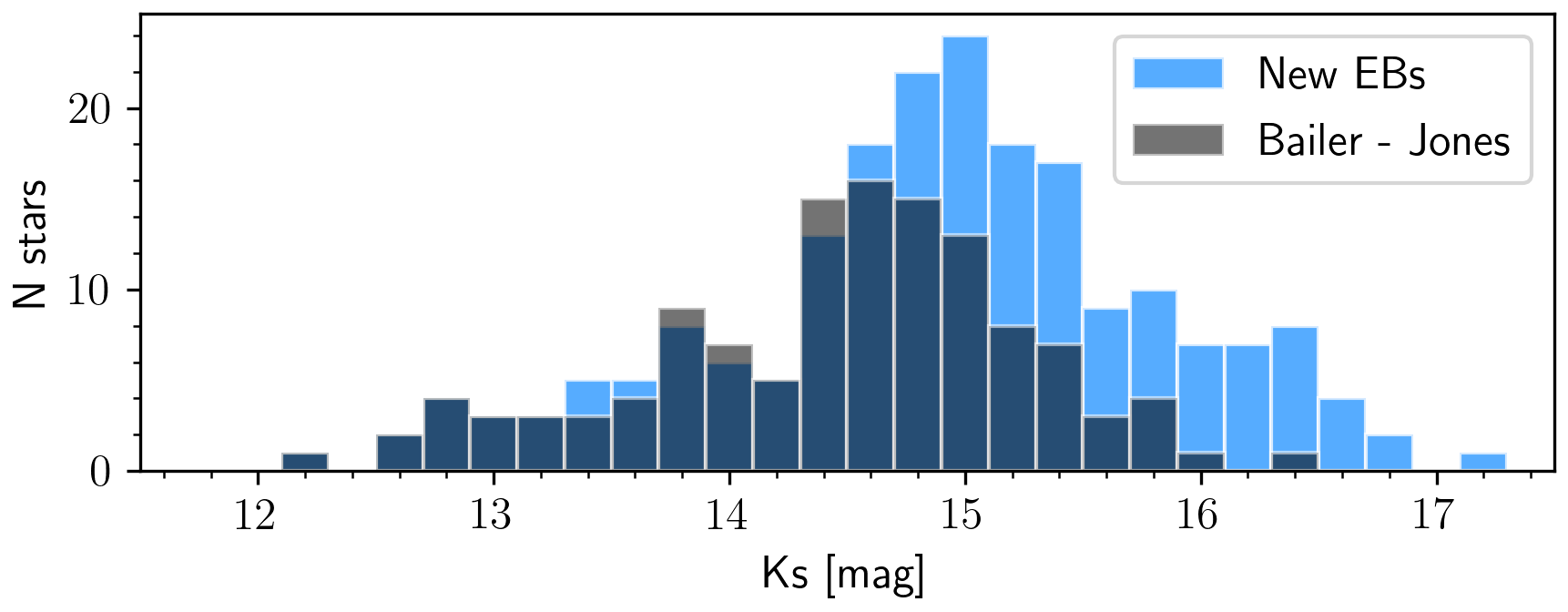}
    \caption{On the top there are Gaia distances by \citet{Bailer-Jones2018} as function of $K_{\rm s}$ magnitude for the new EBs. On the bottom there are the distribution of $K_{\rm s}$ magnitude for the new EBs with Gaia distances and all new EBs.}
    \label{fig:pm_gaia}
\end{figure}

Since astrometric data from Gaia are now available, we cross checked our sample with Gaia DR2 data in order to get optical proper-motions and parallaxes. Only 121 of our EBs have optical counterparts in Gaia, with a large fraction of these ($\sim 23.6\% $) presenting negative parallaxes. Similarly, data for these 121 EBs are present in the Gaia distance catalogue \citep{Bailer-Jones2018} (see Fig. ~\ref{fig:pm_gaia}). While the completeness of Gaia data for our EBs is near $100\%$ up to $K_{\rm s} \sim 14.0$~mag, the Gaia sample is highly incomplete below $K_{\rm s} \sim 15.0$~mag, with only two sources reaching $K_{\rm s} \sim 16.0$~mag. On the other hand, our final sample is about $1$~mag deeper, with a dozen sources below $K_{\rm s} \sim 16.0$~mag and some reaching up to $K_{\rm s} \sim 17$~mag. Distances from \citet{Bailer-Jones2018} suggest that fainter objects in our sample -- those ones without Gaia counterpart -- would be farther and could even belong to the MW bulge ($d>6$ kpc, bottom panel of Fig. ~\ref{fig:pm_gaia}).

We notice that BEAM provides the total extinction integrated along the line of sight, thus the values of $E(J-K_{\rm s})$ as calculated by BEAM are certainly overestimated for nearby objects. While this caveat would favour the use of 3-D maps to yield more accurate extinction values \citep[e.g.][]{2014A&A...566A.120S}, the absence of Gaia distances for a large fraction of our targets prevents us from providing homogeneous 3-D extinctions for all sources in our sample. Values of $A_{K\rm s}$ and $E(J-K_{\rm s})$ as calculated by BEAM for a $2' \times 2'$ area around the position of all new EBs, and assuming the \citet{1989ApJ...345..245C} extinction law, are included in Table A1 in Appendix A.

\begin{table*}
	\centering
	\caption{Most reddened EBs.}
	\label{tab:most_red}
	\begin{tabular}{lccccccc} 
		\hline
ID        & RA         &  DEC     & $K_{\rm s}$  &  $(J-K_{\rm s})$  & $A_{K{\rm s}}$ & $E(J-K_{\rm s})$ &  PM            \\
          & [deg]      &  [deg]   & [mag]  & [mag]       & [mag]    & [mag]     &  [mas y$^{-1}$] \\
\hline
b295-357 & 272.66428 & $-$25.59524 & 13.55$\pm$0.01 & 1.92$\pm$0.03 (\#3) & 0.24$\pm$0.06 & 0.35$\pm$0.08 & 6.48$\pm$0.29 \\
b308-283 & 270.97090 & $-$25.96540 & 15.60$\pm$0.10 & 1.80$\pm$0.18 (\#1) & 1.00$\pm$0.14 & 1.45$\pm$0.21 & 4.63$\pm$2.25 \\
b310-208 & 272.41620 & $-$23.91395 & 14.56$\pm$0.03 & 1.84$\pm$0.05 (\#2) & 0.90$\pm$0.17 & 1.31$\pm$0.25 & 3.35$\pm$0.44 \\
b310-098  & 272.57413 & $-$24.13545 & 15.31$\pm$0.06 & 2.66$\pm$0.16 (\#6) & 1.42$\pm$0.38 & 2.06$\pm$0.55 & 6.69$\pm$0.76 \\
b310-048  & 272.06197 & $-$23.41134 & 15.61$\pm$0.08 & 2.42$\pm$0.18 (\#4) & 0.98$\pm$0.18 & 1.43$\pm$0.26 & 0.49$\pm$1.11 \\ 
b310-276 & 272.55892 & $-$23.56622 & 15.89$\pm$0.10 & 2.60$\pm$0.25 (\#5) & 1.11$\pm$0.21 & 1.61$\pm$0.31 & 2.27$\pm$0.78 \\
\hline
	\end{tabular}
\end{table*}

\begin{figure*}
	\includegraphics[width=\columnwidth]{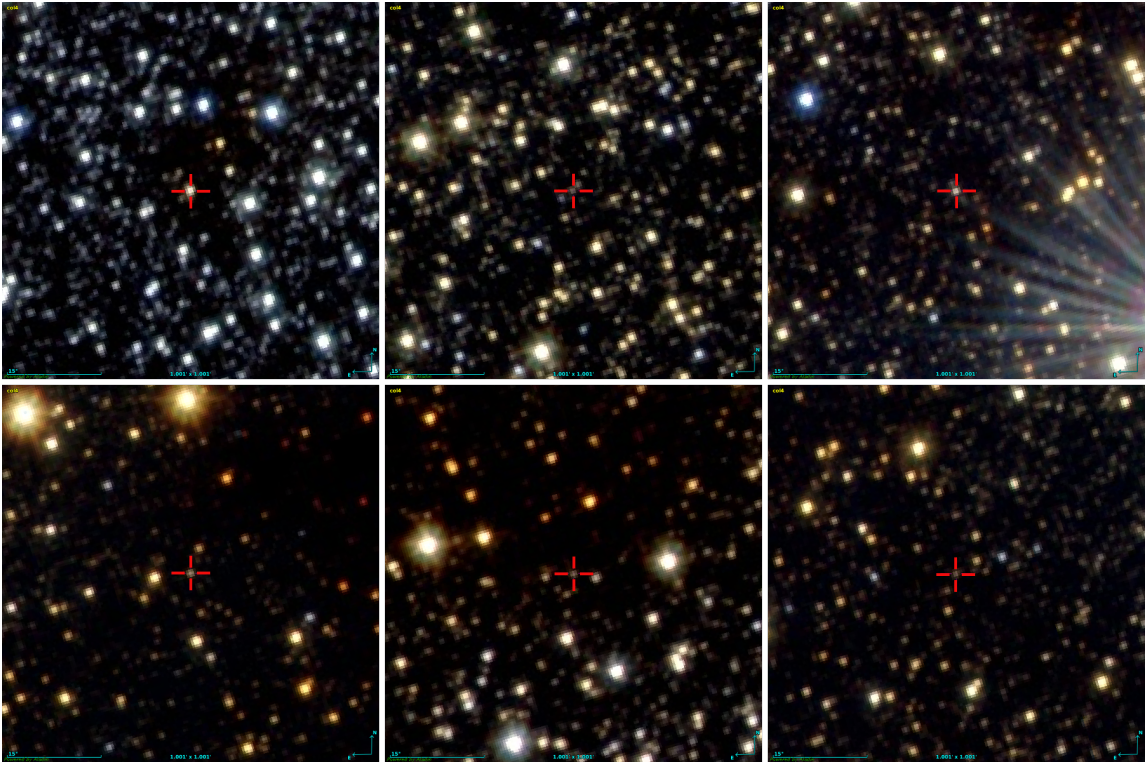}
    \caption{VVV images of the most reddened EBs in our sample. Top-panels: b295$-$357, b308$-$283 and b310$-$208. Bottom-panels: b310$-$098, b310$-$048 and b310$-$276. All fields are $1' \times 1$' in size centred at the target position and oriented in Galactic coordinates.}
    \label{fig:most_red}
\end{figure*}

\subsection{Most reddened EBs}

From the CMD of new EBs (see Fig. \ref{fig:cmd}), six objects presenting $(J-K_{s})>1.8$ mag  caught our attention, since these are the most reddened EBs in our sample. As shown in the spatial distribution of Fig. \ref{fig:fov}, four of these objects are concentrated towards a dark region in tile b310, while one source is in tile b308 and another in tile b295. Table \ref{tab:most_red} summarises their parameters, including the mean reddening and extinction. Fig. \ref{fig:most_red} presents VVV near-IR images for these sources, showing how affected by reddening are the surrounding areas. Except for b295$-$357, all sources are towards the dark regions extending for almost the entire 1' $\times$ 1' images, especially those located in the VVV tile b310.

In applying the total extinction integrated along the line of sight -- as provided by BEAM -- three objects stand out because of their very red intrinsic colours: b310$-$048 with $(J-K_{\rm s})_0=0.99\pm0.32$ mag, b310$-$276 with $(J-K_{\rm s})_0=0.99\pm0.40$ mag and b295$-$357 with $(J-K_{\rm s})_0=1.57\pm0.09$ mag. These colours may be overestimated for nearby objects since we are applying an integrated value of extinction.

A search at the position of b310$-$048 in the VizieR Catalogue Access Tool\footnote{\url{https://vizier.u-strasbg.fr/viz-bin/VizieR}} resulted in a few measurements in the near-IR as well as in longer wavelengths. Near-IR magnitudes from UKIDSS-GPS DR6 \citep{Lucas_2008} are similar to the VVV magnitudes, with $J=18.155 \pm 0.096$ and $K=15.607 \pm 0.044$. In the mid-IR b310$-$048 was observed at the $3.6$ and $4.5$ passbands by GLIMPSE \citep{IPAC_2009}, with $[3.6] =14.953 \pm 0.164$ mag and $[4.5]= 14.924 \pm 0.274$ mag, resulting in $(K_{\rm s}-4.5)=0.69\pm0.29$ mag. The total proper motion as measured by the VVV Infrared Astrometric Catalogue \citep{Smith_2018} is $PM = 0.49 \pm 1.11$ mas yr$^{-1}$. The low PM value combined with the intrinsic magnitude ($K_{s0}=14.63\pm0.20$ mag) and IR colour suggest that b310$-$048 may be a distant source, even belonging to the MW bulge in the background. 

Catalogue search at the position of b310$-$276 resulted in a single observation by UKIDSS-GPS DR6, with $K =15.927 \pm 0.035$, which is consistent with the VVV magnitude ($K_{\rm s} = 15.89\pm0.10$ mag). VIRAC PM is $PM = 2.27 \pm 0.78$ mas yr$^{-1}$, as expected for foreground disk source. b310$-$276 is the faintest object among the most reddened sample, with  $K_{\rm s0}=14.78\pm0.23$ mag. That may explain the absence of previous measurements in the optical and even at longer wavelengths.

b295$-$357 is the brighter source among the red sample, with $K_{s0}=13.31\pm0.06$ mag. This facilitates observations in optical wavelengths, even for an intrinsically red source. b295$-$357 is in the Initial Gaia Source List \citep[IGSL,][]{Smart_2014} with $BJ=24.650 \pm 0.123$ mag, $RF=20.197 \pm 0.123$ mag and $G=20.271 \pm 0.123$ mag. Interestingly, the object is not present in Gaia DR2, thus Gaia parallaxes (or even distances) are not available. Also in optical wavelengths, Pan-STARRS DR1 \citep{Chambers_2017} magnitudes are $i=21.1970 \pm 0.2994$ mag, $z=19.8645 \pm 0.0786$ and $y=18.4993 \pm 0.4420$. Proper-motions from VIRAC indicates a nearby disk object, with $PM = 6.48 \pm 0.29$ mas yr$^{-1}$.

\section{Conclusions} \label{sec:conclusion}

Searches for pulsating variables in the inner MW using VVV data are contributing also to the discovery and study of eclipsing binaries. Variability surveys in the optical such as OGLE and MOA are limited to the cleaner regions towards the bulge as the example of the Baade's window, partially covered by the present observations. Our analysis shows that VVV can pierce through the most obscure regions, thus becoming a useful tool for the search for variable sources in the innermost MW.

Our search for periodic variables was performed with algorithms based on Lomb-Scargle and Phase Dispersion Minimization methods. Using an almost purely automated classification we select a sample of asymmetric EBs by comparing the depths of the eclipses valleys, complemented with a visual search on the symmetric light curves. 

We have classified 212 new bonafide EBs, which present high-quality light curves selected over a conservative threshold ($R^2~{\rm index} > 0.6$). Our new variables are in the period range of $0.1-1.0$ day with an accuracy of $\gtrsim 98\%$ based on the comparison with previously known objects. This period range is interesting because it is shared by the RR Lyrae variables. Our search was performed towards a region of the MW bulge with a total area of 13.4 sq. deg. within $1.68^{\rm o}<l<7.53^{\rm o}$ and $-3.73^{\rm o}<b<-1.44^{\rm o}$, corresponding to the VVV tiles b293 to b296 and b307 to b310. The new objects are not evenly distributed across the area but concentrate towards a small corner of the region, where the extinction is severe, including objects with high reddening up to $E(J-K_{\rm s})=2$ mag. Those are mostly distributed along the foreground disk and with a few also in MW bulge. The absence of optical information from Gaia DR2 for objects brighter than $K_{\rm s} = 15$ mag supports this argument. 

It would be interesting to explore these binaries through determination of orbital parameters using simulated models, as done by PHOEBE, in order to obtain their distances and map the innermost MW bulge. This would be especially important for light curves below $R^2~{\rm index} = 0.6$, which are present in large numbers and are not included in this first analysis. While these regions were used in a first search for EBs in the VVV bulge because of the large number of epochs observed, similar analysis in the innermost MW bulge ($|b|<2$) will certainly contribute to the discovery of a large number of new EBs towards the most obscured regions of the Galaxy.

\section*{Acknowledgements}

We gratefully acknowledge the use of data from the ESO Public Surveys VVV and VVVX (programme IDs 179.B-2002 and 198.B-2004) taken with the VISTA telescope, and data products from the Cambridge Astronomical Survey Unit. This research has made use of the VizieR catalogue access tool, CDS, Strasbourg, France (DOI: 10.26093/cds/vizier). The original description of the VizieR service was published in A\&AS 143, 23. We also thank the physics graduate program of the Universidade Federal de Santa Catarina and the Universidade Federal de Mato Grosso/Brazil for the support. R.K.S. and T.S.F. acknowledges support from CNPq/Brazil through project 305902/2019-9.

\section*{Data Availability} \label{data_availability}

The data underlying this article were accessed from the VISTA Science Archive\footnote{\url{http://horus.roe.ac.uk/vsa/}}, Gaia Archive\footnote{\url{https://archives.esac.esa.int/gaia}} and VizieR Catalogue Access Tool\footnote{\url{https://vizier.u-strasbg.fr/viz-bin/VizieR}}. The derived data generated in this research will be shared on reasonable request to the corresponding author. 



\bibliographystyle{mnras}
\bibliography{paper} 




\appendix

\onecolumn
\section{General information about the new Eclipsing Binaries}

\begin{longtable}{lccccccc}
\caption{\label{tab1} General information about the new EBs. Coordinates are J2000 from the VVV catalogues as well as the magnitudes and colours (see Section 2). For the Period determination see Section 3. Extinction and reddening are integrated values from BEAM (see Section 5). A few of $(J-K_{\rm s})$ colours are missing because of the absence of $J$-band measurements for the target in the VVV data.}\\
\hline
\hline

\vspace {2pt}
    ID       & RA        & DEC         & Period     & $K_{\rm s}$  & $(J-K_{\rm s})$    & A$_{K\rm s}$               & $E(J-K_{\rm s})$   \\
             & {[}deg{]} & {[}deg{]}   & {[}days{]} & {[}mag{]}   & {[}mag{]}        & {[}mag{]}         & {[}mag{]}             \\
\hline
\endfirsthead
\caption{continued.} \\
\hline
\hline
\vspace {2pt}
    ID       & RA        & DEC         & Period     & $K_{\rm s}$  & $(J-K_{\rm s})$    & A$_{K\rm s}$               & $E(J-K_{\rm s})$   \\
             & {[}deg{]} & {[}deg{]}   & {[}days{]} & {[}mag{]}   & {[}mag{]}        & {[}mag{]}         & {[}mag{]}             \\
\hline
\endhead
\hline
\endfoot
b293-068 & 270.69749 & $-$27.88619 & 0.87771 & 15.08$\pm$0.08 &        ---   & 0.25$\pm$0.08 & 0.36$\pm$0.11 \\
b293-197 & 270.58404 & $-$28.30166 & 0.32344 & 15.33$\pm$0.10 & 0.78$\pm$0.14 & 0.17$\pm$0.06 & 0.25$\pm$0.09 \\
b293-213 & 270.93521 & $-$27.68489 & 0.34970 & 14.79$\pm$0.06 & 0.52$\pm$0.08 & 0.19$\pm$0.06 & 0.28$\pm$0.09 \\
b293-230 & 270.33605 & $-$28.31211 & 0.50451 & 14.25$\pm$0.04 & 0.43$\pm$0.05 & 0.22$\pm$0.06 & 0.31$\pm$0.08 \\
b294-032 & 271.25918 & $-$27.69694 & 0.28796 & 15.73$\pm$0.12 &         ---   & 0.18$\pm$0.06 & 0.25$\pm$0.09 \\
b294-083 & 272.12432 & $-$26.91506 & 0.26201 & 15.87$\pm$0.13 & 0.68$\pm$0.18 & 0.22$\pm$0.05 & 0.32$\pm$0.08 \\
b294-210 & 271.83934 & $-$26.45052 & 0.33198 & 14.62$\pm$0.04 & 0.65$\pm$0.06 & 0.27$\pm$0.06 & 0.40$\pm$0.08 \\
b294-257 & 271.92701 & $-$26.93424 & 0.28540 & 12.87$\pm$0.01 & 0.52$\pm$0.01 & 0.17$\pm$0.06 & 0.24$\pm$0.08 \\
b294-323 & 272.02560 & $-$26.66868 & 0.33982 & 14.47$\pm$0.04 & 0.57$\pm$0.05 & 0.19$\pm$0.06 & 0.28$\pm$0.09 \\
b294-341 & 271.42900 & $-$27.52955 & 0.44800 & 10.21$\pm$0.01 & 1.47$\pm$0.01 & 0.14$\pm$0.06 & 0.20$\pm$0.08 \\
b295-048 & 272.53750 & $-$26.72117 & 0.34618 & 14.20$\pm$0.03 & 0.62$\pm$0.03 & 0.27$\pm$0.06 & 0.39$\pm$0.09 \\
b295-056 & 272.17810 & $-$26.12049 & 0.35804 & 15.06$\pm$0.06 & 0.64$\pm$0.07 & 0.27$\pm$0.06 & 0.39$\pm$0.09 \\
b295-086 & 272.29401 & $-$25.21632 & 0.23878 & 15.17$\pm$0.06 & 0.98$\pm$0.09 & 0.37$\pm$0.06 & 0.54$\pm$0.09 \\
b295-089 & 272.35788 & $-$25.22808 & 0.49290 & 16.46$\pm$0.19 & 0.69$\pm$0.24 & 0.36$\pm$0.07 & 0.52$\pm$0.10 \\
b295-164 & 272.52134 & $-$25.43947 & 0.26591 & 15.70$\pm$0.10 & 1.02$\pm$0.14 & 0.29$\pm$0.07 & 0.42$\pm$0.10 \\
b295-165 & 272.52175 & $-$25.30507 & 0.38035 & 13.26$\pm$0.01 & 0.58$\pm$0.01 & 0.31$\pm$0.05 & 0.45$\pm$0.08 \\
b295-233 & 273.06103 & $-$25.75921 & 0.37953 & 14.66$\pm$0.04 & 0.59$\pm$0.05 & 0.23$\pm$0.06 & 0.33$\pm$0.08 \\
b295-236 & 272.59556 & $-$25.18032 & 0.48052 & 14.58$\pm$0.04 & 0.60$\pm$0.04 & 0.34$\pm$0.06 & 0.50$\pm$0.09 \\
b295-246 & 272.00415 & $-$26.12192 & 0.32233 & 15.52$\pm$0.08 & 0.88$\pm$0.11 & 0.30$\pm$0.06 & 0.43$\pm$0.09 \\
b295-357 & 272.66428 & $-$25.59524 & 1.26939 & 13.55$\pm$0.01 & 1.92$\pm$0.03 & 0.24$\pm$0.06 & 0.35$\pm$0.08 \\
b295-367 & 272.71354 & $-$25.21800 & 0.76485 & 15.26$\pm$0.07 & 0.86$\pm$0.09 & 0.34$\pm$0.07 & 0.49$\pm$0.09 \\
b295-382 & 272.20019 & $-$26.23889 & 0.48853 & 13.06$\pm$0.01 & 0.71$\pm$0.01 & 0.28$\pm$0.07 & 0.41$\pm$0.10 \\
b296-001 & 273.19827 & $-$24.40434 & 0.82205 & 15.74$\pm$0.09 & 0.67$\pm$0.11 & 0.37$\pm$0.06 & 0.53$\pm$0.09 \\
b296-013 & 273.78551 & $-$24.85457 & 0.37780 & 14.70$\pm$0.04 & 0.60$\pm$0.04 & 0.26$\pm$0.06 & 0.38$\pm$0.09 \\
b296-038 & 273.33338 & $-$25.40009 & 0.30988 & 13.66$\pm$0.02 & 0.92$\pm$0.02 & 0.26$\pm$0.07 & 0.38$\pm$0.10 \\
b296-053 & 273.02364 & $-$24.72467 & 0.63027 & 15.14$\pm$0.05 & 1.13$\pm$0.07 & 0.51$\pm$0.08 & 0.74$\pm$0.11 \\
b296-054 & 273.05257 & $-$24.84411 & 1.02080 & 14.49$\pm$0.03 & 0.93$\pm$0.04 & 0.49$\pm$0.07 & 0.71$\pm$0.10 \\
b296-068 & 273.36750 & $-$24.56633 & 0.33360 & 15.04$\pm$0.05 & 0.89$\pm$0.06 & 0.35$\pm$0.06 & 0.51$\pm$0.08 \\
b296-073 & 272.97420 & $-$23.99401 & 0.41871 & 15.57$\pm$0.08 & 0.98$\pm$0.09 & 0.52$\pm$0.06 & 0.75$\pm$0.09 \\
b296-074 & 272.98742 & $-$24.00418 & 0.79878 & 14.40$\pm$0.03 & 0.90$\pm$0.04 & 0.51$\pm$0.06 & 0.73$\pm$0.09 \\
b296-079 & 273.15945 & $-$24.01199 & 0.42711 & 14.93$\pm$0.05 & 0.70$\pm$0.05 & 0.40$\pm$0.06 & 0.57$\pm$0.08 \\
b296-096 & 272.94026 & $-$25.24318 & 0.31256 & 14.43$\pm$0.03 & 0.72$\pm$0.03 & 0.34$\pm$0.05 & 0.49$\pm$0.08 \\
b296-102 & 272.69428 & $-$24.60363 & 0.55514 & 15.08$\pm$0.05 & 1.21$\pm$0.07 & 0.61$\pm$0.06 & 0.88$\pm$0.09 \\
b296-104 & 272.71714 & $-$24.74295 & 0.31379 & 16.52$\pm$0.18 & 1.18$\pm$0.26 & 0.60$\pm$0.08 & 0.87$\pm$0.11 \\
b296-105 & 272.72435 & $-$24.54299 & 0.26757 & 16.17$\pm$0.08 &       ---      & 0.66$\pm$0.08 & 0.96$\pm$0.11 \\
b296-107 & 272.77490 & $-$24.63202 & 1.23062 & 15.43$\pm$0.07 & 1.03$\pm$0.09 & 0.61$\pm$0.07 & 0.88$\pm$0.10 \\
b296-126 & 273.53693 & $-$24.63802 & 0.70080 & 15.26$\pm$0.06 & 0.47$\pm$0.07 & 0.35$\pm$0.06 & 0.51$\pm$0.09 \\
b296-136 & 273.65229 & $-$24.76521 & 0.38635 & 13.94$\pm$0.02 & 0.58$\pm$0.02 & 0.30$\pm$0.06 & 0.43$\pm$0.09 \\
b296-140 & 273.19902 & $-$24.08412 & 0.44149 & 14.46$\pm$0.03 & 0.87$\pm$0.04 & 0.41$\pm$0.06 & 0.60$\pm$0.09 \\
b296-143 & 273.26567 & $-$24.01389 & 0.56999 & 15.15$\pm$0.05 & 0.78$\pm$0.07 & 0.37$\pm$0.06 & 0.54$\pm$0.09 \\
b296-144 & 273.30336 & $-$24.18651 & 0.48720 & 15.04$\pm$0.05 & 0.76$\pm$0.06 & 0.38$\pm$0.06 & 0.55$\pm$0.09 \\
b296-148 & 273.32764 & $-$24.00191 & 0.31351 & 14.73$\pm$0.04 & 0.76$\pm$0.05 & 0.40$\pm$0.06 & 0.58$\pm$0.08 \\
b296-150 & 273.70845 & $-$24.38100 & 0.37390 & 14.02$\pm$0.02 & 0.62$\pm$0.02 & 0.32$\pm$0.06 & 0.47$\pm$0.09 \\
b296-171 & 272.88793 & $-$24.82733 & 0.34927 & 16.37$\pm$0.08 &       ---     & 0.52$\pm$0.08 & 0.75$\pm$0.12 \\
b296-188 & 273.14342 & $-$24.30111 & 0.44073 & 14.17$\pm$0.02 & 0.73$\pm$0.03 & 0.44$\pm$0.06 & 0.63$\pm$0.09 \\
b296-198 & 273.66035 & $-$24.47442 & 0.81386 & 15.19$\pm$0.06 & 0.38$\pm$0.06 & 0.29$\pm$0.07 & 0.42$\pm$0.09 \\
b296-212 & 273.39616 & $-$24.00441 & 1.13562 & 14.84$\pm$0.04 & 0.78$\pm$0.05 & 0.39$\pm$0.06 & 0.57$\pm$0.09 \\
b296-221 & 273.87539 & $-$24.12909 & 0.58697 & 15.97$\pm$0.11 & 0.63$\pm$0.13 & 0.23$\pm$0.06 & 0.33$\pm$0.09 \\
b296-226 & 272.67514 & $-$24.88635 & 0.96214 & 15.36$\pm$0.06 & 0.85$\pm$0.08 & 0.44$\pm$0.06 & 0.64$\pm$0.09 \\
b296-259 & 273.57804 & $-$24.40300 & 0.31328 & 14.41$\pm$0.03 &      ---      & 0.32$\pm$0.06 & 0.47$\pm$0.09 \\
b296-260 & 273.14999 & $-$23.78162 & 0.42184 & 15.32$\pm$0.06 & 0.79$\pm$0.08 & 0.46$\pm$0.06 & 0.67$\pm$0.09 \\
b296-269 & 273.66650 & $-$24.11593 & 0.42048 & 14.81$\pm$0.04 & 0.63$\pm$0.05 & 0.32$\pm$0.05 & 0.47$\pm$0.08 \\
b296-271 & 273.68461 & $-$24.04397 & 0.43888 & 12.94$\pm$0.01 & 0.50$\pm$0.01 & 0.29$\pm$0.06 & 0.42$\pm$0.09 \\
b296-288 & 272.70889 & $-$24.44439 & 0.51323 & 14.78$\pm$0.04 & 1.35$\pm$0.05 & 0.65$\pm$0.08 & 0.95$\pm$0.12 \\
b296-290 & 272.73070 & $-$24.54204 & 0.35104 & 16.23$\pm$0.10 &      ---      & 0.66$\pm$0.08 & 0.95$\pm$0.11 \\
b296-299 & 273.36553 & $-$24.70013 & 0.31649 & 16.20$\pm$0.14 & 1.06$\pm$0.19 & 0.37$\pm$0.07 & 0.53$\pm$0.10 \\
b296-308 & 273.00527 & $-$24.16388 & 0.50311 & 13.95$\pm$0.02 & 0.77$\pm$0.02 & 0.42$\pm$0.06 & 0.61$\pm$0.09 \\
b296-314 & 273.33610 & $-$24.42611 & 0.43563 & 14.71$\pm$0.04 & 0.73$\pm$0.04 & 0.32$\pm$0.06 & 0.46$\pm$0.08 \\
b296-318 & 273.38122 & $-$24.25897 & 0.81691 & 14.10$\pm$0.02 & 0.77$\pm$0.03 & 0.44$\pm$0.07 & 0.63$\pm$0.10 \\
b296-322 & 273.89733 & $-$24.61064 & 0.39623 & 15.27$\pm$0.06 & 0.58$\pm$0.07 & 0.29$\pm$0.07 & 0.42$\pm$0.09 \\
b296-327 & 273.49745 & $-$23.99904 & 0.96471 & 16.08$\pm$0.12 & 0.85$\pm$0.16 & 0.40$\pm$0.07 & 0.58$\pm$0.11 \\
b296-329 & 273.50572 & $-$23.91920 & 0.46645 & 14.77$\pm$0.04 & 0.77$\pm$0.05 & 0.34$\pm$0.06 & 0.50$\pm$0.08 \\
b296-332 & 273.57079 & $-$24.00254 & 0.87482 & 15.04$\pm$0.04 &      ---      & 0.33$\pm$0.07 & 0.49$\pm$0.10 \\
b296-357 & 273.17266 & $-$24.61449 & 0.46069 & 15.34$\pm$0.06 & 0.73$\pm$0.08 & 0.41$\pm$0.06 & 0.60$\pm$0.09 \\
b307-059 & 269.63877 & $-$27.18479 & 1.19996 & 14.38$\pm$0.06 & 0.95$\pm$0.08 & 0.54$\pm$0.08 & 0.78$\pm$0.11 \\
b307-060 & 269.63895 & $-$27.34655 & 0.56235 & 13.72$\pm$0.03 & 0.67$\pm$0.04 & 0.42$\pm$0.07 & 0.61$\pm$0.10 \\
b307-067 & 268.99456 & $-$28.29733 & 1.31980 & 14.74$\pm$0.08 & 0.96$\pm$0.11 & 0.51$\pm$0.09 & 0.74$\pm$0.13 \\
b307-071 & 269.12021 & $-$28.24432 & 0.75337 & 12.78$\pm$0.01 & 1.09$\pm$0.02 & 0.40$\pm$0.06 & 0.58$\pm$0.09 \\
b307-100 & 269.47018 & $-$27.68354 & 0.93058 & 13.14$\pm$0.02 & 0.64$\pm$0.02 & 0.46$\pm$0.11 & 0.67$\pm$0.16 \\
b307-118 & 269.81080 & $-$27.31490 & 0.48582 & 15.62$\pm$0.16 & 1.12$\pm$0.25 & 0.40$\pm$0.07 & 0.58$\pm$0.11 \\
b307-129 & 269.12544 & $-$28.30950 & 0.36539 & 16.06$\pm$0.24 & 0.83$\pm$0.32 & 0.43$\pm$0.07 & 0.63$\pm$0.10 \\
b307-141 & 269.45303 & $-$27.98380 & 1.80978 & 14.88$\pm$0.08 & 0.94$\pm$0.12 & 0.50$\pm$0.07 & 0.72$\pm$0.10 \\
b307-144 & 270.02511 & $-$28.29644 & 0.33714 & 15.16$\pm$0.11 & 0.68$\pm$0.14 & 0.27$\pm$0.06 & 0.38$\pm$0.09 \\
b307-168 & 270.01708 & $-$27.20718 & 0.29102 & 16.56$\pm$0.11 &      ---      & 0.40$\pm$0.07 & 0.58$\pm$0.10 \\
b308-028 & 270.82003 & $-$26.19713 & 0.32433 & 14.63$\pm$0.05 & 0.90$\pm$0.06 & 0.48$\pm$0.10 & 0.70$\pm$0.14 \\
b308-035 & 271.35636 & $-$26.45456 & 0.37921 & 16.42$\pm$0.24 & 0.51$\pm$0.28 & 0.23$\pm$0.06 & 0.33$\pm$0.08 \\
b308-062 & 270.82195 & $-$26.67581 & 1.37136 & 12.92$\pm$0.01 & 0.74$\pm$0.01 & 0.44$\pm$0.09 & 0.64$\pm$0.13 \\
b308-066 & 270.47261 & $-$25.99344 & 0.52096 & 12.56$\pm$0.01 & 0.97$\pm$0.01 & 0.81$\pm$0.12 & 1.18$\pm$0.17 \\
b308-074 & 270.99118 & $-$26.22163 & 0.74501 & 15.40$\pm$0.10 & 0.80$\pm$0.13 & 0.52$\pm$0.10 & 0.76$\pm$0.14 \\
b308-084 & 270.31727 & $-$27.19938 & 0.23843 & 13.84$\pm$0.02 & 1.15$\pm$0.03 & 0.38$\pm$0.06 & 0.55$\pm$0.09 \\
b308-101 & 270.09841 & $-$26.72024 & 0.43869 & 16.52$\pm$0.26 & 1.28$\pm$0.39 & 0.67$\pm$0.11 & 0.97$\pm$0.16 \\
b308-122 & 270.49451 & $-$26.47302 & 0.39197 & 14.51$\pm$0.04 & 0.68$\pm$0.05 & 0.52$\pm$0.09 & 0.75$\pm$0.13 \\
b308-138 & 270.72858 & $-$26.15104 & 0.36061 & 15.39$\pm$0.10 & 0.97$\pm$0.13 & 0.58$\pm$0.09 & 0.84$\pm$0.14 \\
b308-143 & 271.20520 & $-$26.47363 & 0.40729 & 15.10$\pm$0.08 & 0.77$\pm$0.09 & 0.37$\pm$0.07 & 0.54$\pm$0.10 \\
b308-147 & 269.96969 & $-$27.10360 & 1.19542 & 13.30$\pm$0.02 & 0.60$\pm$0.02 & 0.41$\pm$0.07 & 0.60$\pm$0.10 \\
b308-159 & 270.30116 & $-$26.72215 & 1.07390 & 14.75$\pm$0.06 & 0.77$\pm$0.07 & 0.48$\pm$0.08 & 0.70$\pm$0.12 \\
b308-180 & 271.10248 & $-$26.46762 & 0.43416 & 14.23$\pm$0.04 &      ---      & 0.40$\pm$0.06 & 0.58$\pm$0.09 \\
b308-185 & 270.80151 & $-$25.93967 & 0.97017 & 15.66$\pm$0.13 & 1.50$\pm$0.21 & 0.91$\pm$0.13 & 1.32$\pm$0.19 \\
b308-186 & 270.81901 & $-$26.01989 & 0.37126 & 14.78$\pm$0.06 & 0.73$\pm$0.07 & 0.66$\pm$0.17 & 0.96$\pm$0.24 \\
b308-193 & 270.16433 & $-$26.85860 & 0.40151 & 16.44$\pm$0.10 &     ---      & 0.59$\pm$0.09 & 0.86$\pm$0.13 \\
b308-203 & 270.29770 & $-$26.57464 & 1.04726 & 14.53$\pm$0.05 & 0.95$\pm$0.06 & 0.57$\pm$0.10 & 0.83$\pm$0.14 \\
b308-237 & 270.98943 & $-$26.14896 & 1.60287 & 16.45$\pm$0.25 & 1.20$\pm$0.35 & 0.67$\pm$0.13 & 0.97$\pm$0.18 \\
b308-251 & 270.26195 & $-$26.46642 & 0.32402 & 13.63$\pm$0.02 & 1.39$\pm$0.03 & 0.59$\pm$0.08 & 0.85$\pm$0.12 \\
b308-264 & 270.48932 & $-$26.16261 & 0.58169 & 15.86$\pm$0.15 & 1.69$\pm$0.27 & 0.77$\pm$0.15 & 1.12$\pm$0.22 \\
b308-274 & 270.81701 & $-$26.03205 & 0.44469 & 14.21$\pm$0.03 &      ---      & 0.64$\pm$0.16 & 0.93$\pm$0.23 \\
b308-283 & 270.97090 & $-$25.96540 & 0.93347 & 15.60$\pm$0.10 & 1.80$\pm$0.18 & 1.00$\pm$0.14 & 1.45$\pm$0.21 \\
b309-012 & 271.98722 & $-$25.57935 & 1.07181 & 14.60$\pm$0.05 &      ---       & 0.29$\pm$0.06 & 0.42$\pm$0.09 \\
b309-021 & 271.71717 & $-$24.91486 & 0.64874 & 15.92$\pm$0.09 &      ---       & 0.73$\pm$0.10 & 1.05$\pm$0.14 \\
b309-029 & 270.99411 & $-$25.78442 & 0.27549 & 14.82$\pm$0.05 & 0.91$\pm$0.06 & 0.76$\pm$0.12 & 1.11$\pm$0.17 \\
b309-030 & 270.99965 & $-$25.81846 & 0.34061 & 16.20$\pm$0.16 & 1.24$\pm$0.22 & 0.79$\pm$0.13 & 1.14$\pm$0.18 \\
b309-032 & 271.04769 & $-$25.88579 & 0.71651 & 15.73$\pm$0.08 &     ---      & 0.78$\pm$0.11 & 1.13$\pm$0.16 \\
b309-034 & 271.05712 & $-$25.93318 & 1.09313 & 13.46$\pm$0.01 & 0.95$\pm$0.02 & 0.75$\pm$0.11 & 1.09$\pm$0.16 \\
b309-047 & 271.71062 & $-$25.83656 & 0.47046 & 15.09$\pm$0.06 & 0.52$\pm$0.07 & 0.31$\pm$0.06 & 0.44$\pm$0.09 \\
b309-049 & 271.48944 & $-$25.34697 & 0.43173 & 14.93$\pm$0.05 & 0.92$\pm$0.07 & 0.52$\pm$0.08 & 0.76$\pm$0.11 \\
b309-058 & 271.66125 & $-$25.37029 & 0.22079 & 15.18$\pm$0.07 & 0.89$\pm$0.08 & 0.46$\pm$0.08 & 0.67$\pm$0.12 \\
b309-059 & 271.66318 & $-$25.29830 & 0.66787 & 15.23$\pm$0.07 & 1.01$\pm$0.09 & 0.58$\pm$0.08 & 0.84$\pm$0.11 \\
b309-060 & 271.16571 & $-$24.75942 & 0.66785 & 15.07$\pm$0.06 & 1.44$\pm$0.09 & 0.84$\pm$0.18 & 1.22$\pm$0.27 \\
b309-062 & 271.22517 & $-$24.71269 & 0.77280 & 15.69$\pm$0.11 & 1.43$\pm$0.16 & 0.94$\pm$0.16 & 1.36$\pm$0.23 \\
b309-063 & 271.24541 & $-$24.81083 & 0.34119 & 15.33$\pm$0.04 &     ---       & 0.74$\pm$0.10 & 1.07$\pm$0.15 \\
b309-065 & 271.29355 & $-$24.80976 & 0.48893 & 15.95$\pm$0.14 & 1.73$\pm$0.23 & 0.76$\pm$0.10 & 1.10$\pm$0.15 \\
b309-076 & 270.64910 & $-$25.63030 & 1.11064 & 14.83$\pm$0.05 & 0.95$\pm$0.06 & 0.66$\pm$0.10 & 0.95$\pm$0.15 \\
b309-077 & 270.64956 & $-$25.77970 & 0.65385 & 13.70$\pm$0.02 & 0.80$\pm$0.02 & 0.68$\pm$0.11 & 0.98$\pm$0.16 \\
b309-083 & 270.73358 & $-$25.75736 & 1.93084 & 15.09$\pm$0.06 & 1.42$\pm$0.09 & 0.75$\pm$0.14 & 1.10$\pm$0.20 \\
b309-111 & 271.31057 & $-$25.22080 & 0.28569 & 14.40$\pm$0.03 & 0.97$\pm$0.04 & 0.66$\pm$0.10 & 0.95$\pm$0.14 \\
b309-124 & 271.48244 & $-$24.88831 & 0.52238 & 16.20$\pm$0.17 & 1.05$\pm$0.21 & 0.66$\pm$0.09 & 0.95$\pm$0.13 \\
b309-151 & 271.48823 & $-$25.73753 & 0.62374 & 14.33$\pm$0.03 & 0.57$\pm$0.03 & 0.36$\pm$0.07 & 0.52$\pm$0.10 \\
b309-180 & 271.65415 & $-$24.68120 & 0.89205 & 15.09$\pm$0.06 & 1.01$\pm$0.08 & 0.61$\pm$0.08 & 0.88$\pm$0.11 \\
b309-199 & 270.99241 & $-$25.68553 & 0.43012 & 15.26$\pm$0.07 & 0.89$\pm$0.09 & 0.68$\pm$0.12 & 0.99$\pm$0.17 \\
b309-220 & 271.60416 & $-$25.21537 & 1.19643 & 14.71$\pm$0.04 & 0.92$\pm$0.05 & 0.55$\pm$0.09 & 0.80$\pm$0.14 \\
b309-222 & 271.63386 & $-$25.21951 & 0.58302 & 13.71$\pm$0.02 & 1.00$\pm$0.02 & 0.54$\pm$0.09 & 0.78$\pm$0.13 \\
b309-254 & 270.86155 & $-$25.27081 & 0.49617 & 14.69$\pm$0.04 & 0.67$\pm$0.05 & 0.59$\pm$0.10 & 0.86$\pm$0.15 \\
b309-261 & 271.06641 & $-$25.21966 & 0.70271 & 13.32$\pm$0.01 & 0.57$\pm$0.02 & 0.62$\pm$0.10 & 0.89$\pm$0.14 \\
b309-264 & 271.49109 & $-$25.46975 & 0.78882 & 14.92$\pm$0.05 & 0.69$\pm$0.06 & 0.41$\pm$0.07 & 0.59$\pm$0.10 \\
b309-271 & 271.13461 & $-$24.96094 & 0.79692 & 16.52$\pm$0.22 & 1.23$\pm$0.29 & 0.70$\pm$0.10 & 1.01$\pm$0.15 \\
b309-276 & 271.23263 & $-$24.88162 & 0.87911 & 14.83$\pm$0.05 & 1.63$\pm$0.08 & 0.74$\pm$0.11 & 1.08$\pm$0.16 \\
b309-291 & 271.63388 & $-$24.75779 & 0.27060 & 14.46$\pm$0.04 & 1.04$\pm$0.05 & 0.65$\pm$0.08 & 0.94$\pm$0.12 \\
b309-299 & 272.17650 & $-$25.01973 & 0.54746 & 13.84$\pm$0.02 & 0.58$\pm$0.02 & 0.35$\pm$0.07 & 0.51$\pm$0.10 \\
b309-307 & 271.16478 & $-$25.82092 & 0.61851 & 13.89$\pm$0.02 & 0.82$\pm$0.03 & 0.60$\pm$0.09 & 0.87$\pm$0.13 \\
b310-007 & 272.38362 & $-$23.62231 & 0.38198 & 13.95$\pm$0.02 & 1.09$\pm$0.02 & 0.78$\pm$0.11 & 1.13$\pm$0.15 \\
b310-008 & 272.43407 & $-$23.66464 & 0.60860 & 15.01$\pm$0.05 & 1.51$\pm$0.06 & 1.02$\pm$0.16 & 1.48$\pm$0.24 \\
b310-009 & 272.44574 & $-$23.75309 & 1.34384 & 15.34$\pm$0.06 & 1.27$\pm$0.08 & 0.75$\pm$0.10 & 1.08$\pm$0.15 \\
b310-010 & 272.44859 & $-$23.71423 & 0.47355 & 16.65$\pm$0.19 & 1.42$\pm$0.25 & 0.85$\pm$0.13 & 1.24$\pm$0.19 \\
b310-011 & 272.46990 & $-$23.71720 & 0.63615 & 15.30$\pm$0.06 & 1.63$\pm$0.08 & 0.86$\pm$0.09 & 1.25$\pm$0.14 \\
b310-017 & 272.92776 & $-$23.88293 & 0.41608 & 13.46$\pm$0.01 & 1.17$\pm$0.02 & 0.63$\pm$0.09 & 0.91$\pm$0.13 \\
b310-020 & 273.02523 & $-$23.88621 & 1.49323 & 14.57$\pm$0.03 & 0.90$\pm$0.04 & 0.48$\pm$0.07 & 0.70$\pm$0.10 \\
b310-022 & 271.72343 & $-$24.61193 & 0.48942 & 14.92$\pm$0.04 & 1.04$\pm$0.05 & 0.63$\pm$0.08 & 0.91$\pm$0.11 \\
b310-024 & 271.77573 & $-$24.61223 & 1.93093 & 13.17$\pm$0.01 & 1.07$\pm$0.01 & 0.67$\pm$0.08 & 0.98$\pm$0.12 \\
b310-025 & 271.81858 & $-$24.71331 & 1.11226 & 14.64$\pm$0.03 & 0.95$\pm$0.04 & 0.65$\pm$0.09 & 0.94$\pm$0.12 \\
b310-026 & 271.85471 & $-$24.67036 & 0.92983 & 14.95$\pm$0.04 & 0.86$\pm$0.05 & 0.71$\pm$0.09 & 1.03$\pm$0.13 \\
b310-033 & 271.99767 & $-$24.20779 & 0.39605 & 15.70$\pm$0.08 & 1.49$\pm$0.11 & 0.76$\pm$0.10 & 1.10$\pm$0.14 \\
b310-039 & 272.50582 & $-$24.54691 & 0.96111 & 16.33$\pm$0.14 & 1.08$\pm$0.17 & 0.65$\pm$0.07 & 0.94$\pm$0.10 \\
b310-046 & 272.02794 & $-$23.52108 & 0.91415 & 15.30$\pm$0.06 & 1.33$\pm$0.08 & 0.71$\pm$0.09 & 1.03$\pm$0.13 \\
b310-047 & 272.06186 & $-$23.48966 & 0.40711 & 14.41$\pm$0.03 & 0.88$\pm$0.03 & 0.67$\pm$0.13 & 0.97$\pm$0.19 \\
b310-048 & 272.06197 & $-$23.41134 & 0.29931 & 15.61$\pm$0.08 & 2.42$\pm$0.18 & 0.98$\pm$0.18 & 1.43$\pm$0.26 \\
b310-049 & 272.07375 & $-$23.59777 & 0.44475 & 15.53$\pm$0.08 & 1.18$\pm$0.09 & 0.64$\pm$0.09 & 0.93$\pm$0.14 \\
b310-051 & 272.50246 & $-$23.71959 & 1.12346 & 14.88$\pm$0.04 & 1.50$\pm$0.06 & 0.83$\pm$0.06 & 1.21$\pm$0.09 \\
b310-055 & 271.39138 & $-$24.50722 & 0.89603 & 14.92$\pm$0.06 & 1.33$\pm$0.08 & 0.94$\pm$0.15 & 1.36$\pm$0.22 \\
b310-060 & 271.97559 & $-$24.60855 & 0.30423 & 14.54$\pm$0.03 & 0.85$\pm$0.03 & 0.66$\pm$0.07 & 0.96$\pm$0.10 \\
b310-061 & 271.97690 & $-$24.62266 & 0.70213 & 13.56$\pm$0.01 & 0.75$\pm$0.02 & 0.64$\pm$0.07 & 0.93$\pm$0.10 \\
b310-070 & 272.14019 & $-$24.33077 & 0.45167 & 17.11$\pm$0.28 &     ---      & 0.73$\pm$0.08 & 1.06$\pm$0.11 \\
b310-074 & 272.22346 & $-$24.38133 & 0.64560 & 15.30$\pm$0.06 & 1.34$\pm$0.07 & 0.77$\pm$0.11 & 1.12$\pm$0.16 \\
b310-086 & 272.01026 & $-$23.99337 & 0.72492 & 15.01$\pm$0.04 & 1.51$\pm$0.06 & 0.73$\pm$0.11 & 1.06$\pm$0.17 \\
b310-091 & 272.04845 & $-$23.96629 & 0.73538 & 16.02$\pm$0.11 & 1.53$\pm$0.15 & 0.78$\pm$0.14 & 1.13$\pm$0.20 \\
b310-093 & 272.07679 & $-$23.97599 & 0.34928 & 16.31$\pm$0.14 & 1.56$\pm$0.20 & 0.86$\pm$0.11 & 1.25$\pm$0.15 \\
b310-097 & 272.52979 & $-$24.12898 & 1.10625 & 15.07$\pm$0.05 &      ---      & 1.51$\pm$0.24 & 2.19$\pm$0.34 \\
b310-098 & 272.57413 & $-$24.13545 & 0.36339 & 15.31$\pm$0.06 & 2.66$\pm$0.16 & 1.42$\pm$0.38 & 2.06$\pm$0.55 \\
b310-102 & 272.14258 & $-$23.60754 & 0.43526 & 13.78$\pm$0.02 & 0.90$\pm$0.02 & 0.62$\pm$0.07 & 0.90$\pm$0.11 \\
b310-104 & 272.17879 & $-$23.59753 & 0.33719 & 16.11$\pm$0.12 & 1.24$\pm$0.15 & 0.64$\pm$0.07 & 0.93$\pm$0.10 \\
b310-109 & 272.64449 & $-$23.88306 & 0.50433 & 16.44$\pm$0.10 &     ---      & 0.66$\pm$0.07 & 0.96$\pm$0.10 \\
b310-112 & 272.68429 & $-$23.75018 & 0.59593 & 15.06$\pm$0.05 & 1.15$\pm$0.06 & 0.72$\pm$0.09 & 1.04$\pm$0.13 \\
b310-113 & 272.69290 & $-$23.76985 & 0.73943 & 15.30$\pm$0.06 & 1.38$\pm$0.08 & 0.70$\pm$0.10 & 1.02$\pm$0.15 \\
b310-127 & 272.14361 & $-$24.75106 & 0.29727 & 16.50$\pm$0.17 & 1.04$\pm$0.20 & 0.55$\pm$0.08 & 0.80$\pm$0.11 \\
b310-134 & 271.79446 & $-$24.23587 & 1.73263 & 15.80$\pm$0.09 & 1.50$\pm$0.12 & 0.62$\pm$0.07 & 0.90$\pm$0.10 \\
b310-138 & 272.33343 & $-$24.53006 & 0.45790 & 14.69$\pm$0.03 & 0.88$\pm$0.04 & 0.65$\pm$0.06 & 0.95$\pm$0.09 \\
b310-144 & 272.04449 & $-$23.66959 & 0.60364 & 14.02$\pm$0.02 & 0.79$\pm$0.02 & 0.62$\pm$0.09 & 0.90$\pm$0.13 \\
b310-149 & 272.59705 & $-$24.01349 & 0.49356 & 15.41$\pm$0.06 & 1.43$\pm$0.08 & 0.70$\pm$0.11 & 1.02$\pm$0.16 \\
b310-150 & 272.61655 & $-$24.00538 & 0.54332 & 15.86$\pm$0.10 & 1.15$\pm$0.11 & 0.60$\pm$0.08 & 0.88$\pm$0.12 \\
b310-157 & 272.66347 & $-$24.10511 & 0.30278 & 14.69$\pm$0.03 & 1.41$\pm$0.04 & 0.70$\pm$0.08 & 1.02$\pm$0.11 \\
b310-159 & 272.70838 & $-$23.93108 & 0.63461 & 16.42$\pm$0.16 & 1.20$\pm$0.19 & 0.57$\pm$0.09 & 0.82$\pm$0.13 \\
b310-163 & 272.33195 & $-$23.41885 & 0.39106 & 15.22$\pm$0.06 & 1.67$\pm$0.08 & 0.87$\pm$0.15 & 1.27$\pm$0.21 \\
b310-164 & 272.37878 & $-$23.36820 & 0.38659 & 15.07$\pm$0.05 & 1.13$\pm$0.06 & 0.71$\pm$0.10 & 1.02$\pm$0.14 \\
b310-166 & 272.84439 & $-$23.73463 & 0.72155 & 16.14$\pm$0.12 & 0.92$\pm$0.14 & 0.58$\pm$0.07 & 0.84$\pm$0.11 \\
b310-167 & 272.85310 & $-$23.75120 & 0.60052 & 14.62$\pm$0.03 & 1.38$\pm$0.04 & 0.56$\pm$0.07 & 0.81$\pm$0.10 \\
b310-168 & 272.85758 & $-$23.68945 & 0.43174 & 14.48$\pm$0.03 & 0.96$\pm$0.03 & 0.55$\pm$0.08 & 0.80$\pm$0.11 \\
b310-171 & 271.64964 & $-$24.30483 & 0.52603 & 12.66$\pm$0.01 & 0.65$\pm$0.01 & 0.73$\pm$0.08 & 1.06$\pm$0.11 \\
b310-172 & 271.76222 & $-$24.47585 & 0.36705 & 15.96$\pm$0.10 & 1.18$\pm$0.13 & 0.75$\pm$0.09 & 1.09$\pm$0.13 \\
b310-173 & 271.77672 & $-$24.30480 & 1.42832 & 12.20$\pm$0.01 & 0.56$\pm$0.01 & 0.62$\pm$0.07 & 0.89$\pm$0.10 \\
b310-180 & 272.29788 & $-$24.61166 & 0.56158 & 15.48$\pm$0.07 & 0.93$\pm$0.08 & 0.54$\pm$0.06 & 0.78$\pm$0.09 \\
b310-189 & 271.91825 & $-$23.99116 & 0.35548 & 14.65$\pm$0.03 & 0.88$\pm$0.04 & 0.60$\pm$0.08 & 0.88$\pm$0.11 \\
b310-197 & 271.99208 & $-$24.06596 & 0.53031 & 12.78$\pm$0.01 & 0.84$\pm$0.01 & 0.67$\pm$0.09 & 0.98$\pm$0.13 \\
b310-201 & 272.38183 & $-$24.37079 & 1.41593 & 15.30$\pm$0.06 & 1.72$\pm$0.09 & 0.95$\pm$0.12 & 1.37$\pm$0.17 \\
b310-203 & 272.42277 & $-$24.40036 & 0.53192 & 14.98$\pm$0.04 & 1.39$\pm$0.06 & 0.90$\pm$0.09 & 1.30$\pm$0.13 \\
b310-204 & 272.42456 & $-$24.38547 & 0.60585 & 14.45$\pm$0.03 &      ---      & 0.91$\pm$0.09 & 1.33$\pm$0.13 \\
b310-208 & 272.41620 & $-$23.91395 & 0.81730 & 14.56$\pm$0.03 & 1.84$\pm$0.05 & 0.90$\pm$0.17 & 1.31$\pm$0.25 \\
b310-211 & 272.47629 & $-$23.98593 & 0.40617 & 14.59$\pm$0.03 & 1.13$\pm$0.04 & 0.74$\pm$0.10 & 1.07$\pm$0.15 \\
b310-222 & 272.12075 & $-$23.40349 & 0.32763 & 14.21$\pm$0.02 & 1.46$\pm$0.03 & 0.85$\pm$0.15 & 1.23$\pm$0.22 \\
b310-224 & 272.12657 & $-$23.40442 & 1.02265 & 14.73$\pm$0.04 & 1.23$\pm$0.04 & 0.82$\pm$0.14 & 1.19$\pm$0.20 \\
b310-225 & 272.12712 & $-$23.24393 & 0.37959 & 16.27$\pm$0.08 &      ---      & 0.65$\pm$0.10 & 0.94$\pm$0.15 \\
b310-226 & 272.17160 & $-$23.33962 & 0.76096 & 15.25$\pm$0.06 & 1.24$\pm$0.07 & 0.70$\pm$0.13 & 1.01$\pm$0.19 \\
b310-228 & 272.62808 & $-$23.62366 & 0.70909 & 16.30$\pm$0.14 & 1.29$\pm$0.18 & 0.79$\pm$0.10 & 1.14$\pm$0.15 \\
b310-232 & 272.69033 & $-$23.66245 & 1.31307 & 15.45$\pm$0.07 & 1.30$\pm$0.08 & 0.69$\pm$0.09 & 1.01$\pm$0.14 \\
b310-233 & 272.74789 & $-$23.56757 & 0.37734 & 15.80$\pm$0.09 & 1.03$\pm$0.11 & 0.60$\pm$0.08 & 0.87$\pm$0.12 \\
b310-234 & 271.48519 & $-$24.29371 & 0.79456 & 15.24$\pm$0.06 & 1.36$\pm$0.07 & 0.98$\pm$0.13 & 1.42$\pm$0.19 \\
b310-251 & 271.72356 & $-$23.98917 & 0.92717 & 15.16$\pm$0.04 &      ---      & 0.65$\pm$0.10 & 0.94$\pm$0.15 \\
b310-254 & 272.19637 & $-$24.20073 & 0.25645 & 16.70$\pm$0.20 & 1.26$\pm$0.24 & 0.71$\pm$0.10 & 1.02$\pm$0.15 \\
b310-255 & 272.30065 & $-$24.29852 & 0.32617 & 16.00$\pm$0.11 & 1.04$\pm$0.13 & 0.66$\pm$0.06 & 0.95$\pm$0.09 \\
b310-256 & 272.37017 & $-$24.19480 & 0.35332 & 14.86$\pm$0.04 & 1.54$\pm$0.05 & 0.81$\pm$0.10 & 1.17$\pm$0.14 \\
b310-257 & 271.84754 & $-$23.65731 & 0.98346 & 15.24$\pm$0.06 & 1.63$\pm$0.08 & 0.86$\pm$0.14 & 1.25$\pm$0.21 \\
b310-259 & 271.89703 & $-$23.65713 & 0.32019 & 13.83$\pm$0.02 & 0.93$\pm$0.02 & 0.71$\pm$0.09 & 1.03$\pm$0.13 \\
b310-260 & 271.91711 & $-$23.58875 & 0.77766 & 14.92$\pm$0.04 & 1.30$\pm$0.05 & 0.75$\pm$0.10 & 1.08$\pm$0.14 \\
b310-267 & 272.04267 & $-$23.59449 & 0.36370 & 15.37$\pm$0.06 & 1.08$\pm$0.07 & 0.64$\pm$0.10 & 0.93$\pm$0.14 \\
b310-271 & 272.78700 & $-$23.99384 & 0.51687 & 14.57$\pm$0.03 & 1.02$\pm$0.04 & 0.62$\pm$0.08 & 0.89$\pm$0.12 \\
b310-272 & 272.39979 & $-$23.52183 & 0.64650 & 15.14$\pm$0.05 & 1.11$\pm$0.06 & 0.88$\pm$0.25 & 1.28$\pm$0.37 \\
b310-273 & 272.44286 & $-$23.56450 & 0.50687 & 16.23$\pm$0.13 & 1.46$\pm$0.18 & 0.82$\pm$0.13 & 1.19$\pm$0.19 \\
b310-274 & 272.48251 & $-$23.58641 & 0.30832 & 16.16$\pm$0.13 & 1.35$\pm$0.16 & 0.96$\pm$0.23 & 1.40$\pm$0.33 \\
b310-275 & 272.54935 & $-$23.47133 & 0.99023 & 12.71$\pm$0.01 & 1.28$\pm$0.01 & 0.79$\pm$0.15 & 1.15$\pm$0.21 \\
b310-276 & 272.55892 & $-$23.56622 & 0.47570 & 15.89$\pm$0.10 & 2.60$\pm$0.25 & 1.11$\pm$0.21 & 1.61$\pm$0.31 \\
b310-282 & 273.02303 & $-$23.82756 & 0.26484 & 13.77$\pm$0.02 & 0.95$\pm$0.02 & 0.50$\pm$0.07 & 0.72$\pm$0.10 \\
b310-303 & 272.03554 & $-$24.19751 & 0.37541 & 16.86$\pm$0.23 & 0.87$\pm$0.26 & 0.74$\pm$0.09 & 1.08$\pm$0.13 \\
b310-304 & 272.07250 & $-$24.15921 & 1.39400 & 14.77$\pm$0.04 & 1.22$\pm$0.04 & 0.83$\pm$0.09 & 1.21$\pm$0.12 \\
b310-306 & 272.12369 & $-$24.11055 & 1.26208 & 14.94$\pm$0.04 & 1.53$\pm$0.06 & 0.91$\pm$0.11 & 1.32$\pm$0.16 \\
b310-308 & 272.13474 & $-$24.11196 & 0.87431 & 15.55$\pm$0.07 & 1.34$\pm$0.09 & 0.85$\pm$0.12 & 1.24$\pm$0.18 \\
\end{longtable}




\bsp	
\label{lastpage}
\end{document}